\documentclass[prx,aps,twocolumn,showpacs,floatfix]{revtex4-1}
\usepackage{amsfonts}
\usepackage{amssymb}
\usepackage{hyperref}
\usepackage{graphicx}
\usepackage{dcolumn}
\usepackage{bm,amsmath,verbatim}
\usepackage{mathrsfs}
\usepackage{color}
\usepackage{sidecap}
\allowdisplaybreaks
\usepackage{soul}
\usepackage{dsfont}
\usepackage{amsmath,amssymb,amsthm}
\usepackage{subfigure}
\usepackage{amsfonts,amssymb}
\usepackage{float}

\hypersetup{colorlinks,
linkcolor=blue,          citecolor=blue,        filecolor=blue,      urlcolor=blue           }

\def\be{\begin{equation}}
\def\ee{\end{equation}}
\def\bea{\begin{eqnarray}}
\def\eea{\end{eqnarray}}
\def\bse{\begin{subequations}}
\def\ese{\end{subequations}}

\def\be{\begin{eqnarray}}
\def\ee{\end{eqnarray}}

\begin{document}

\title{Higher-order Topological Anderson Insulators}
\author{Yan-Bin Yang$^{1}$}
\thanks{These authors contribute equally to this work.}
\author{Kai Li$^{1}$}
\thanks{These authors contribute equally to this work.}
\author{L.-M. Duan$^{1}$}
\author{Yong Xu$^{1,2}$}
\email{yongxuphy@tsinghua.edu.cn}
\affiliation{$^{1}$Center for Quantum Information, IIIS, Tsinghua University, Beijing 100084, People's Republic of China}
\affiliation{$^{2}$Shanghai Qi Zhi Institute, Shanghai 200030, People's Republic of China}

\begin{abstract}
We study disorder effects in a two-dimensional system with chiral symmetry and find
that disorder can induce a quadrupole topological insulating phase
(a higher-order topological phase with quadrupole moments)
from a topologically trivial phase. Their topological properties manifest in a
topological invariant defined based on effective boundary Hamiltonians,
the quadrupole moment and zero-energy corner modes.
We find gapped and gapless topological phases and a Griffiths regime.
In the gapless topological phase, all the states
are localized, while in the Griffiths regime, the states at zero energy become multifractal.
We further apply the self-consistent Born approximation to show that the induced topological phase
arises from disorder renormalized masses. We finally introduce a practical experimental scheme with
topolectrical circuits where the predicted topological phenomena can be observed by impedance measurements. Our work opens the door to studying
higher-order topological Anderson insulators and their localization properties.
\end{abstract}
\maketitle

\section{Introduction}

Traditional topological phases usually feature the bulk-boundary correspondence that $(n-1)$-dimensional gapless
boundary states exist for an $n$-dimensional topological system. Recently, topological phases have been generalized
to the case where there exist $(n-m)$-dimensional (instead of $n-1$) gapless boundary states with $1<m\le n$
for an $n$-dimensional system~\cite{Taylor2017Science,Fritz2012PRL,ZhangFan2013PRL}.
In the past few years, the higher-order topological phenomena have drawn tremendous attention, and
various higher-order topological states have been discovered~\cite{Taylor2017Science,Fritz2012PRL,ZhangFan2013PRL,Slager2015PRB,FangChen2017PRL,
Brouwer2017PRL,Bernevig2018SciAdv,Brouwer2019PRX,
Roy2019PRB,SYang2019PRL,Vincent2020PRL,Haiping2020PRL,Yanbin2020PRR,Tiwari2020PRL}, such as quadrupole topological
phases with zero-energy corner modes~\cite{Taylor2017Science} and its type-II cousin~\cite{Yanbin2020PRR} and second-order topological insulators with
chiral hinge modes~\cite{Bernevig2018SciAdv}.
It has also been shown that higher-order topological insulators (HOTIs) are robust against weak disorder~\cite{Hatsugai2019PRB,Jiang2019CPB,Franca2019PRB,CALi2020PRB,Agarwala2020PRR,XRWang2020}.

Disorder plays an important role in quantum transport, such as Anderson localization and
metal-insulator transitions~\cite{Evers2008RMP}. In the context of first-order topological phases,
it has been shown that they are usually stable against weak symmetry preserving disorder.
But disorder is not always detrimental to first-order topological phases.
Ref.~\cite{Sheng2009PRL} theoretically predicted that disorder can drive a topological
phase transition from a metallic trivial phase to a quantum spin Hall insulator; topological insulators induced by disorder
are called topological Anderson insulators (TAIs)~\cite{Sheng2009PRL,Beenakker2009PRL}. Since their discovery, there
has been great interest and advancement in the study of TAIs~\cite{Jiang2009PRB,Franz2010PRL,Altland2014PRL,Prodan2014PRL,Rafael2015PRL,SZhang2017PRL}.
In addition, disorder can drive a transition from a Weyl semimetal to a 3D quantum anomalous Hall state~\cite{Xie2015PRL}. Remarkably, the TAI has been experimentally observed in a photonic waveguide array~\cite{Szameit2018Nat} and disordered cold atomic wire~\cite{Gadway2018Science}.

Disorder, topology and symmetry are closely connected, which can be
seen from classification theories. For example,
random matrix theories are classified based on three internal symmetries, explaining universal transport
properties of disordered physical systems~\cite{Altland1997PRB,Beenakker1997RMP,HaakeBook}. Similarly, the classification
of topological phases is made according to these internal symmetries~\cite{Ludwig2010NJP}.
Among these symmetries, chiral symmetry plays an important role in
disordered systems and many peculiar properties have been found, such as
the divergence of density of states (DOS) and localization length at energy $E=0$~\cite{Dyson,Cohen1976PRB,Eggarter1978PRB,Eilmes1998EPJB,Brouwer1998PRL}.
In 2D, first-order topological phases are not allowed in a system with only chiral symmetry.
Yet, it has been reported that a second-order topological phase can exist
in a 2D system with chiral symmetry~\cite{Okugawa2019PRB,Qibo2020PRB} and thus provides an ideal platform
to study the interplay between disorder and topology.

Here we study the interplay between disorder and higher-order topology in a 2D system
with chiral symmetry. We prove that the quantization of the quadrupole moment is maintained
by chiral symmetry irrespective of crystalline symmetries, indicating that the
quadrupole topological insulator can exist in a system with chiral symmetry without the requirement of
any crystalline symmetry. This also gives us an opportunity to
explore the effects of off-diagonal disorder respecting
chiral symmetry. We theoretically predict the existence of a disorder induced
HOTI [dubbed higher-order topological Anderson insulator (HOTAI)]
with zero-energy corner modes, which arises through the localization-delocalization-localization phase transition.
We further apply the self-consistent Born approximation (SCBA) to show that the induced phase
appears due to the disorder renormalized masses. Besides,
we find gapped and gapless HOTAIs and a Griffiths regime.
In the gapless regime, all the states are localized, while in
the Griffiths regime, the states at zero energy become multifractal.
In addition, we study the disorder effects on a HOTI and show the
existence of gapped and gapless topological phases and a Griffiths regime.
Finally, we propose an experimental scheme using topolectrical circuits to realize and detect
the HOTAI.

\section{Model Hamiltonian}
\label{sec2}

\begin{figure}[t]
  \includegraphics[width=3.5in]{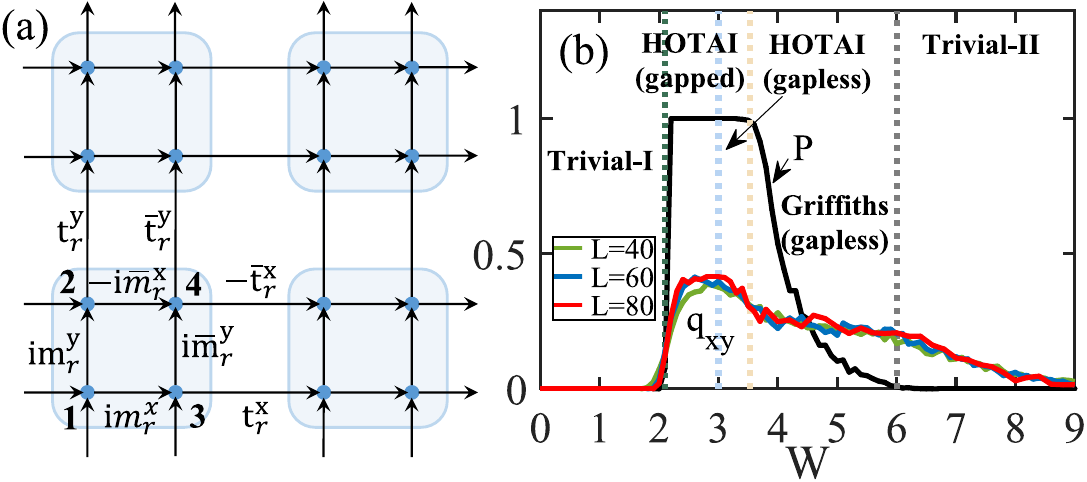}
  \caption{(Color online) (a) Schematics of our model (\ref{Hmodel}).
  (b) The phase diagram with respect to the disorder strength $W$ mapped out based on the topological invariant $P$,
  the quadrupole moments $q_{xy}$, and the bulk energy gap shown in Fig.~\ref{fig2}(a).
  We observe distinct phases including
  gapped/gapless HOTAI, Griffiths phase, and trivial I/II phases, separated by vertical dashed lines.
  Here $m_x=m_y=1.1$.}
\label{fig1}
\end{figure}

We start by considering the following higher-order Hamiltonian
\begin{equation}
\hat{H}=\sum_{\bf r} [ \hat{c}^\dagger_{\bf r}h_0\hat{c}_{\bf r}
+( \hat{c}^\dagger_{\bf r}h_x\hat{c}_{{\bf r}+{\bf e}_x}
 +\hat{c}^\dagger_{\bf r}h_y\hat{c}_{{\bf r}+{\bf e}_y}+H.c. ) ],
\label{Hmodel}
\end{equation}
where $\hat{c}^\dagger_{\bf r}=\left(
                                   \begin{array}{cccc}
                                     \hat{c}^\dagger_{{\bf r} 1} & \hat{c}^\dagger_{{\bf r} 2} & \hat{c}^\dagger_{{\bf r} 3} &
                                     \hat{c}^\dagger_{{\bf r} 4} \\
                                   \end{array}
                                 \right)
$ with $\hat{c}^\dagger_{{\bf r} \nu}$ ($\hat{c}_{{\bf r} \nu}$) being a creation (annihilation) operator
at the $\nu$th site in a unit cell described by ${\bf r}=(x,y)$ with
$x$ and $y$ being integers (suppose that the lattice constants are equal to one), ${\bf e}_x=(1,0)$
and ${\bf e}_y=(0,1)$.
Here
\begin{equation}
h_0=\left(
  \begin{array}{cccc}
    0 & -im_{{\bf r}}^y & -im_{{\bf r}}^x & 0 \\
    im_{{\bf r}}^y & 0 & 0 & i\bar{m}_{{\bf r}}^x \\
    im_{{\bf r}}^x & 0 & 0 & -i\bar{m}_{{\bf r}}^y \\
    0 & -i\bar{m}_{{\bf r}}^x & i\bar{m}_{{\bf r}}^y & 0 \\
  \end{array}
\right)
\end{equation}
depicts the intra-cell hopping, and
\begin{equation}
\begin{aligned}
&h_x=\left(
      \begin{array}{cccc}
        0 & 0 & 0 & 0 \\
        0 & 0 & 0 & 0 \\
        t_{{\bf r}}^x & 0 & 0 & 0 \\
        0 & -\bar{t}_{{\bf r}}^x & 0 & 0 \\
      \end{array}
    \right)\text{and}~
h_y=\left(
      \begin{array}{cccc}
        0 & 0 & 0 & 0 \\
        t_{{\bf r}}^y & 0 & 0 & 0 \\
        0 & 0 & 0 & 0 \\
        0 & 0 & \bar{t}_{{\bf r}}^y & 0 \\
      \end{array}
    \right)
\end{aligned}
\end{equation}
describe the inter-cell hopping along $x$ and $y$, respectively [also see Fig.~\ref{fig1}(a)
for the hopping parameters].
The system parameters $m_{{\bf r}}^x$, $\bar{m}_{{\bf r}}^x$, $m_{{\bf r}}^y$, $\bar{m}_{{\bf r}}^y$,
$t_{{\bf r}}^x$, $\bar{t}_{{\bf r}}^x$, $t_{{\bf r}}^y$
and $\bar{t}_{{\bf r}}^y$ all take real values.
For simplicity without loss of generality, we take the inter-cell hopping magnitude
as energy units so that
$t_{\bf r}^x=\bar{t}_{\bf r}^x=t_{\bf r}^y=\bar{t}_{\bf r}^y=1$. In this case, $h_x=\sigma_{-}\otimes \sigma_z$
and $h_y=\sigma_0\otimes \sigma_{-}$ with $\sigma_{-}=[0\, 0; 1\, 0]$ and $\sigma_0$ being a $2\times 2$
identity matrix.
For a clean system with $m_{\bf r}^x=\bar{m}_{\bf r}^x=m_{\bf r}^y=\bar{m}_{\bf r}^y=m$, the system respects a
generalized $C_4$ symmetry as detailed in Appendix A.

To show that the Hamiltonian (\ref{Hmodel}) describes a higher-order phase supporting zero-energy corner modes
in the clean case with $m_{\bf r}^x=\bar{m}_{\bf r}^x=m_x$
and $m_{\bf r}^y=\bar{m}_{\bf r}^y=m_y$, we write the Hamiltonian in momentum space as
\begin{equation}
\hat{H}=\sum_{\bf k} \hat{c}_{\bf k}^\dagger H_0({\bf k})\hat{c}_{\bf k}.
\end{equation}
Here
\begin{equation}
H_0({\bf k})=H_x(k_x,m_x)\otimes \sigma_z+\sigma_0 \otimes H_y(k_y,m_y),
\label{HMmom}
\end{equation}
where $H_\nu(k_\nu,m_\nu)=\cos k_\nu \sigma_x+(m_\nu+\sin k_\nu)\sigma_y$ $(\nu=x,y)$ with $\sigma_\nu$
$(\nu=x,y,z)$ being the Pauli matrices and $\sigma_0$ being a $2\times 2$ identity matrix.
To see the presence of zero-energy corner modes in the system,
we recast the Hamiltonian (\ref{HMmom}) to a form in continuous real space by
replacing $\sin k_\nu$ by $-i\partial_\nu$ and $\cos k_\nu$ by $1+\partial_\nu^2 /2$ ($\nu=x,y$)
so that $H_\nu(k_\nu)\rightarrow \bar{H}_\nu$.
Considering semi-infinite boundaries along $x$ and $y$,
if $|u_{x}\rangle$ and $|u_{y}\rangle$ are zero-energy edge modes of $\bar{H}_x$ and $\bar{H}_y$,
respectively, $|u_{x}\rangle\otimes |u_{y}\rangle$
is a zero-energy mode of $\bar{H}_0$ localized at a corner.

Since the system
contains only the nearest-neighbor hopping, it respects
chiral symmetry, i.e., ${\Pi} H {\Pi}^{-1}=-H$, where
$H$ is the first-quantization Hamiltonian and ${\Pi}$ is
a unitary matrix.
But this system breaks the time-reversal symmetry and thus the
particle-hole symmetry, because $h_0$ is complex.
In contrast, if we generalize the Benalcazar-Bernevig-Hughes (BBH) model~\cite{Taylor2017Science} to the disordered
case, it still respects
the time-reversal, particle-hole and chiral symmetries.
However, these two models are connected through a local transformation and thus have similar
topological and localization properties as proved in Appendix B.
The equivalence also tells us that our system supports
zero-energy corner modes and has quantized quadrupole moments~\cite{Cho2018arXiv,Wheeler2018arXiv}
protected by reflection symmetries. But with disorder breaking the reflection symmetry, one may wonder whether
the quadrupole moment is still quantized.
Here we prove the quantization of
the quadrupole moment maintained by chiral symmetry (see Appendix C),
indicating that chiral symmetry can protect a quadrupole topological insulator.
We remark that in three dimensions (3D) chiral symmetry maintains the quantization of the octupole
moment as proved in Appendix C, indicating that chiral symmetry can protect the third-order topological insulator
with zero-energy corner modes in 3D.

To study the disorder effects,
we consider the disorder
in the intra-cell hopping, that is, $m_{\bf r}^\nu=m_\nu+W^{\nu}V_{\bf r}^\nu$
and $\bar{m}_{\bf r}^\nu={m}_\nu+\bar{W}^{\nu}\bar{V}_{\bf r}^\nu$
with $\nu=x,y$, where ${V}_{\bf r}^\nu$ and $\bar{V}_{\bf r}^\nu$ are
uniformly randomly distributed in $[-0.5,0.5]$ without correlation. Here ${W}^{\nu}$
and $\bar{W}^{\nu}$ represent the disorder strength.
For simplicity,
we take ${W}^{x}=\bar{W}^{x}={W}^{y}=\bar{W}^{y}=W$.
Because of the
random character, we perform the average over 200-2000 sample configurations for numerical calculation.

\begin{figure}[t]
  \includegraphics[width=3.5in]{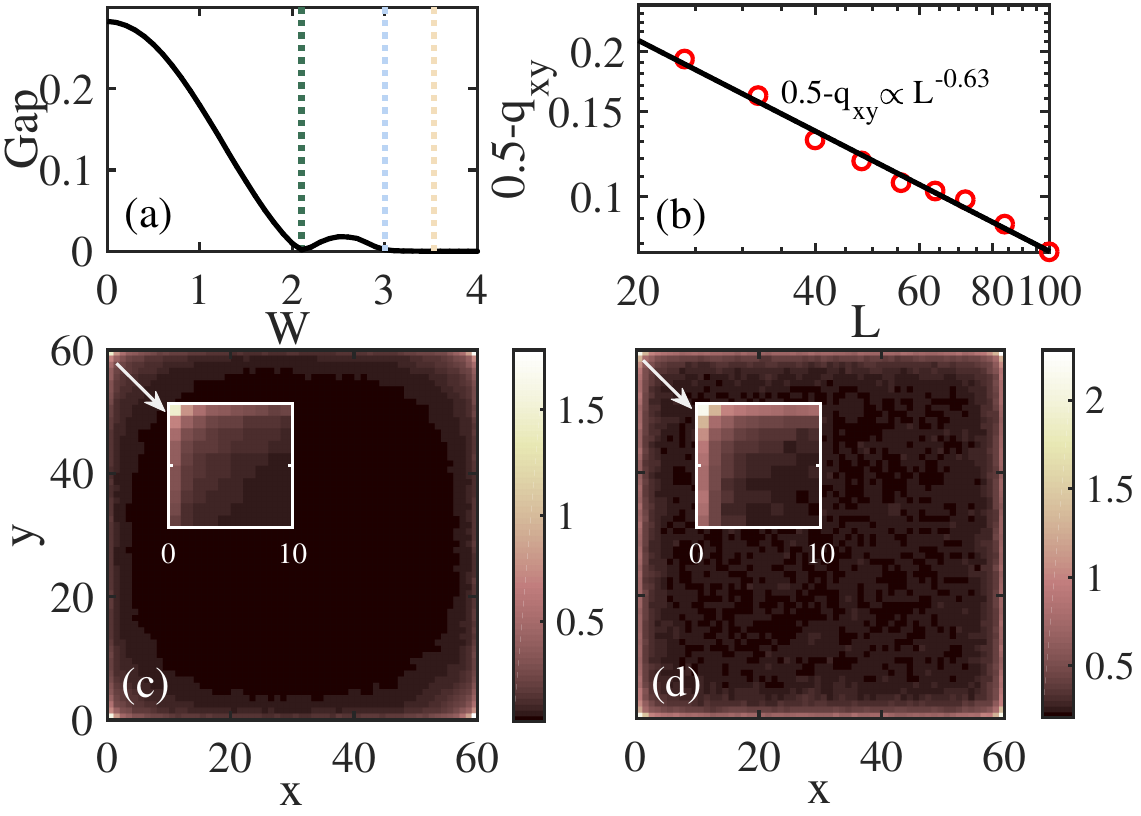}
  \caption{(Color online) (a) The bulk energy gap versus $W$.
  (b) The quadrupole moment (red circles) versus the system size $L$ when $W=2.6$, which
  is fitted by a power-law function plotted as the black line.
  The zero energy LDOS obtained under open boundary conditions
  for (c) $W=2.6$ and (d) $W=3.4$. The zoomed-in view of LDOS around one corner is shown in the insets.
  Here $m_x=m_y=1.1$.}
\label{fig2}
\end{figure}

\section{Higher-order topological Anderson insulators}
\label{sec3}

We map out the phase diagram in Fig.~\ref{fig1}(b),
showing remarkably the presence of a disorder-induced higher-order topological phase transition.
To characterize the phase transition, we evaluate the polarization $p_x$ ($p_y$) of
effective boundary Hamiltonians at a $y$-normal ($x$-normal) boundary at half filling. In one
dimension, the polarization is equivalent to the Berry phase in a translation invariant
system, which can be used as a topological invariant~\cite{Resta}. In fact, the polarization as
a topological invariant can be
evaluated in real space for a system without translational symmetries based on Resta's formula~\cite{Resta,Prodan2014PRL}.
For the quadrupole topological phase, we define
a topological invariant based on $p_x$ and $p_y$ as
\begin{equation}
P=4|p_x p_y|.
\end{equation}
When $P=1$, the system is in a higher-order topologically nontrivial phase, and when $P=0$, it is in
a trivial phase (see Appendix D for its justification for a clean system).

We now generalize it to the disordered case. Specifically, we evaluate the
average polarization of the effective boundary Hamiltonian at the $y$-normal boundary (similarly for $x$-normal one)
by
$
p_x=\frac{1}{N_i}\sum_{n=1}^{N_i} |p_{x,n}|
$,
where $p_{x,n}=\text{Im}\text{log}\langle \Psi_n|e^{2\pi i\hat{x}/L_x}|\Psi_n\rangle/(2\pi)$~\cite{Resta}
with $\hat{x}=\sum_{x}x\hat{n}_x$, $\hat{n}_x$ being the particle number operator at the site $x$,
and $L_x$ being the length of the system along $x$
(we also deduct the atomic positive charge contribution). Here $|\Psi_n\rangle$ is the ground state
at half filling of the boundary Hamiltonian
$H_{n}=-G_{2n}(E=0)^{-1}$ with $G_{2n}$ being the $2n$th boundary
Green's function obtained by~\cite{Dai2008PRB,Oppen2017PRB}
\begin{equation}
G_n=(E-h_n-V_{n-1}G_{n-1}V_{n-1}^\dagger)^{-1},
\end{equation}
where $h_n$ is the Hamiltonian for the $n$th layer and $V_{n-1}$ is the coupling between
the $(n-1)$th and $n$th layer. We note that $p_{x,n}$ is quantized to be either $0$ or $0.5$
for each iteration
since $H_{n}$ also preserves chiral symmetry.
The polarization is evaluated at even steps of Green's function given that
there are two different layers in the clean limit.
In the disordered case,
the intra-cell hopping parts in $h_n$ and $V_{n}$ are randomly generated for each iteration (see Appendix D).
The topological invariant $P$ is finally determined.

In Fig.~\ref{fig1}(b), we plot the topological invariant $P$ as the disorder magnitude
increases. We see that $P$ suddenly jumps to $1$ when $W\approx 2.1$, indicating
the occurrence of a topological phase transition.
$P$ remains quantized to be $1$ until $W>3.5$, where it begins
decreasing continuously. This regime corresponds to the Griffiths phase
where topologically trivial and nontrivial sample configurations coexist (see Appendix E).
When $W> 6$, $P$ vanishes, showing that the system reenters
into a trivial phase.

To further identify that the induced topological phase is a quadrupole topological
phase, we calculate the quadrupole moment, which can be used as a topological invariant
since its quantization is protected
by chiral symmetry (see Appendix C).
Figure~\ref{fig1}(b) shows that the quadrupole
moment qualitatively agrees with the results of $P$. Yet,
conspicuous discrepancy can be observed. The quadrupole moment $q_{xy}$ over many samples is not quantized to $0.5$ in
the regime where $P=1$ [$q_{xy}=0.5$ for most disorder configurations and $q_{xy}=0$ for other configurations
(see Fig.~\ref{figC1} in Appendix C)] and the Griffiths regime is much larger. We attribute this
to the finite-size effects, given that for the quadrupole moment, we can only perform a computation
for a system with its size up to $80$, while to determine $P$, we consider a system with its size up to $500$ and iterations
up to $10^3$. To be more quantitative, we plot $0.5-q_{xy}$ as the system size $L$ increases when $W=2.6$ in Fig.~\ref{fig2}(b),
showing a power law decay and thus suggesting that $q_{xy}$ approaches $0.5$ in the thermodynamic limit.

\begin{figure}[t]
\includegraphics[width=3.4in]{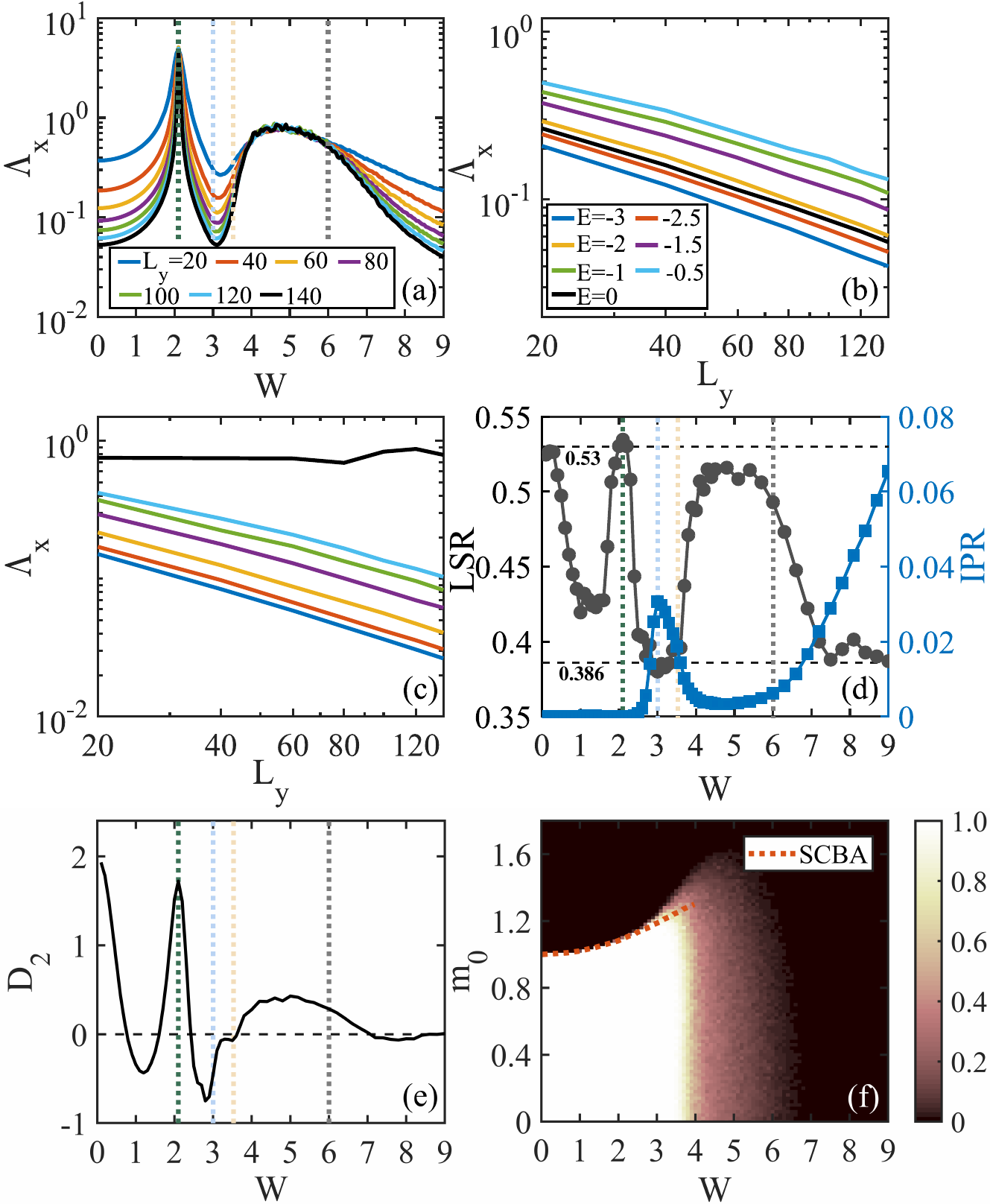}
  \caption{(Color online)
  (a) The normalized localization length $\Lambda_x$ at $E=0$ versus $W$ for several distinct $L_y$.
  The scaling of $\Lambda_x$ at different energies for (b) $W=3.2$ and (c) $W=4.6$.
  (d) The LSR and IPR versus $W$ for the eigenstates around zero energy
  in a system with size $L=500$.
  (e) The fractal dimension $D_2$ with respect to $W$ for the eigenstates
  around zero energy.
  The vertical dashed lines separate different phases.
  (f) The topological invariant $P$ in the $(W,m_0)$ plane with $m_0=m_x=m_y$.
  The red dotted line indicates the topological phase boundary determined by the SCBA.
  In (a-e), $m_x=m_y=1.1$. }
\label{fig3}
\end{figure}

The higher-order topological phase transition occurs as the bulk energy gap closes at $W\approx 2.1$ and reopens,
as shown in Fig.~\ref{fig2}(a). In fact, the transition is associated with the divergence of the localization length at $W\approx 2.1$ [see Fig.~\ref{fig3}(a)].
When $W$ is further increased, the energy gap closes again
and remains closed due to the strong disorder scattering, leading to the gapless HOTAI.
Even in the gapless regime, the topological invariant $P$ can still be quantized
as shown in Fig.~\ref{fig1}(b).
In fact, in this phase, all the states are localized corresponding to
an Anderson insulator (see the following discussion).

To further confirm that the TAI is a higher-order topological state, in Fig.~\ref{fig2}(c-d), we display the local density of states (LDOS)
at $E=0$ for two typical values of $W$ corresponding to
a gapped and gapless topological phase, respectively, clearly showing the presence of
zero-energy states localized at corners. The evidence above definitely suggests the existence
of HOTAIs.

\section{Localization properties}
\label{sec4}

\begin{figure*}[t]
	\includegraphics[width=6.5in]{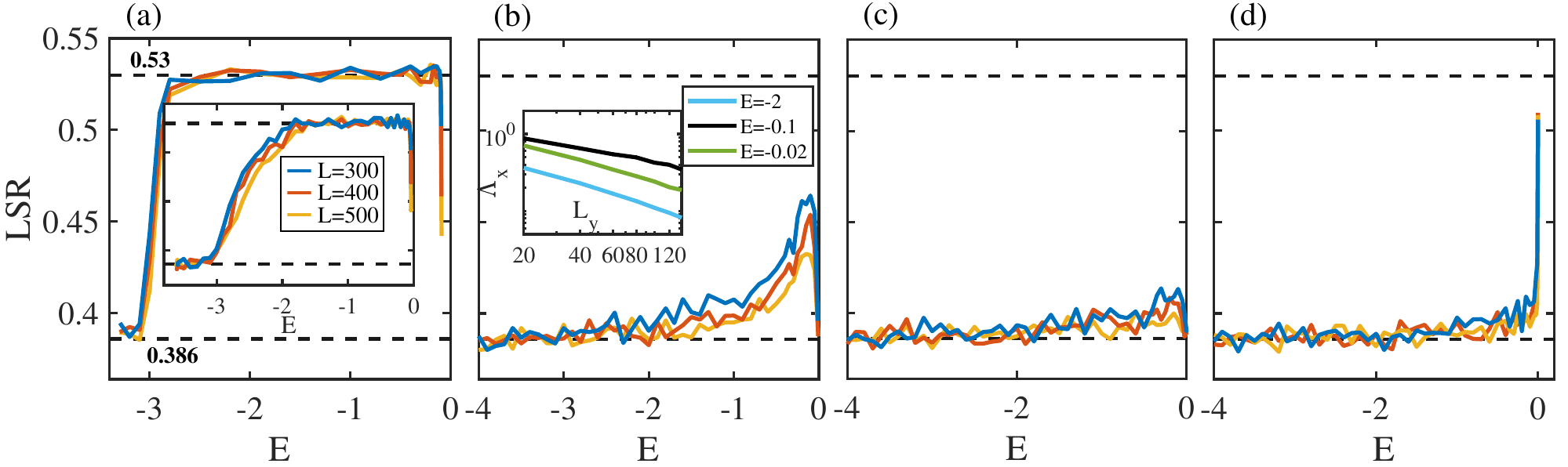}
	\caption{(Color online) The LSR versus energy $E$ for (a) $W=1$, (b) $W=2.8$, (c) $W=3.2$ and (d) $W=4.6$ in the trivial-I, gapped HOTAI, gapless HOTAI and Griffiths phases, respectively.
The blue, red and yellow lines describe results for the system size $L_x=L_y=L=300,400,500$, respectively.
The inset in (a) displays the LSR as a function of $E$ for $W=1.5$.
The inset in (b) shows the normalized localization length $\Lambda_x$ with respect to $L_y$ at different energies.
Here $m_x=m_y=1.1$.
	}
	\label{fig4}
\end{figure*}

We now study the localization properties of energy
bands in different phases by evaluating their localization length, adjacent level-spacing ratio (LSR),
inverse participation ratio (IPR) and fractal dimensions.
The LSR is defined as
\begin{equation}
r(E)=[\frac{1}{N_E-2}\sum_{i}\text{min}(\delta_i,\delta_{i+1})/\text{max}(\delta_i,\delta_{i+1})],
\end{equation}
where $\delta_i=E_i-E_{i-1}$ with $E_i$ being the $i$th eigenenergy sorted in
an ascending order and $\sum_{i}$ denotes the sum over an energy bin around the energy $E$
with $N_E$ energy levels counted.
For localized states, $r\approx 0.386$
corresponding to the Poisson statistics and for extended states
of symmetric real Hamiltonians,
$r\approx 0.53$ corresponding to the Gaussian orthogonal ensemble (GOE)~\cite{Huse2007PRB}.

The localization property can also be characterized by the real space IPR defined as
\begin{equation}
I(E)=[\frac{1}{N_E}\sum_i\sum_{\bf r}(\sum_{\nu=1}^4|\Psi_{E_i,{\bf r}\nu}|^2)^2].
\end{equation}
This quantity evaluates how much a state in an energy bin
around energy $E$ is spatially localized. For an extended state in 2D,
$I\propto 1/L^2$ with $L$ being the size of a system, which goes zero in the thermodynamic limit;
for a state
localized in a single unit cell, it is one. It is well known that at the critical point
between localized and delocalized phases, the state exhibits multifractal behavior with
fractal dimensions $D_2$ defined through $I\propto 1/L^{D_2}$~\cite{Castellani1986}. Clearly, $D_2=2$ and $D_2=0$
indicates that a state is extended and localized, respectively, in the thermodynamic limit;
intermediate values of $D_2$ suggests the multifractal state.

In Fig.~\ref{fig3}(a), we plot the normalized localization length $\Lambda_x=\lambda_x/L_y$ (similarly for $\lambda_y/L_x$) with respect to the disorder strength $W$
at $E=0$ for distinct $L_y$, where $\lambda_\nu$ ($\nu=x,y$) is the localization length along $\nu$ calculated by the
transfer matrix method~\cite{Kramer1983}. In the gapless HOTAI and trivial-II phases, we see the decrease of $\Lambda_x$ as $L_y$ is increased, suggesting
that the states at $E=0$ are localized.
The decline can also be clearly seen in Fig.~\ref{fig3}(b)
where $\Lambda_x$ versus $L_y$ is plotted for $W=3.2$ for distinct energies.
In fact, all states are localized in these two phases as detailed in the following discussion.
This shows that even in the higher-order case, the topology can be
carried by localized bulk states.
Being localized for the states in these regimes is also evidenced by their
relatively large IPR and the LSR approaching $0.386$ [see Fig.~\ref{fig3}(d)].
In these regimes, the fractal dimension $D_2$ becomes negative or approaches zero [see Fig.~\ref{fig3}(e)],
further indicating that the states around zero energy are localized. We note that the negative $D_2$ arises from finite-size effects.
It indicates
that the IPR rises with increasing the system size, suggesting that the states are localized (see Appendix F
for the finite-size analysis).

Figure~\ref{fig3}(a) also demonstrates the existence of a regime (corresponding to the Griffiths regime)
where $\Lambda_x$ at $E=0$ remains almost unchanged as $L_y$ increases, suggesting a multifractal phase
in this regime. The multifractal phase resides between two localized phases, which is very different
from the conventional wisdom that a multifractal phase lives at the critical point between delocalized
and localized phases. In fact, only the states at or very near $E=0$ become multifractal, and
all other states remain localized [see Fig.~\ref{fig3}(c)].
The multifractal properties are also evidenced by the fractal dimension of the states around zero
energy as shown in Fig.~\ref{fig3}(e).

In the gapped regime,
there are trivial-I and gapped HOTAI phases.
In the trivial phase, the states at the band edge around zero energy
exhibit the LSR close to $0.386$ [see Fig.~\ref{fig3}(d)], suggesting the localized
property of these states. The localized property is also evidenced by the negative $D_2$
(in the region around
$W=1$) [see Fig.~\ref{fig3}(e)]. We note that near the phase transition points of $W=0$ and $W=2.1$,
the states exhibit delocalized properties due to the large localization length.
In the gapped HOTAI, Fig.~\ref{fig3}(d) and (e) illustrate that the LSR experiences a drop
from around $0.53$ to $0.386$
and $D_2$ drops from $1.72$ to negative values, suggesting that the states at the band edge
undergo a phase transition from delocalized to localized ones.

The above results indicate that for strong disorder,
all states are localized in the gapless HOTAI and the trivial-II phase. Yet in the Griffiths phase, all states are localized except at $E=0$ where the states become multifractal. For weak disorder, all states can be localized in the topological regime.
In the trivial-I phase, the states at the band edge around zero energy are localized.

In the following, we provide more evidence
on localization properties.
Figure~\ref{fig4} shows the LSR as a function of energy
for five different disorder amplitudes. For small $W$ corresponding to the trivial-I phase
[Fig.~\ref{fig4}(a) and its inset], the LSR remains around $0.53$ except at the lower band edge where
it exhibits a sudden drop towards $0.386$, indicating that the states at the band edge are localized.
But we cannot claim the existence of mobility edges in the trivial-I phase given that it is very possible
that the delocalized behavior is caused by the finite-size effects, which is very difficult to identify
since the localization length is huge for the weak disorder.
In the gapped HOTAI, while we cannot conclusively determine that all states are localized when
$W$ is near the transition point, we show that this occurs when $W$ is larger.
For instance, when $W=2.8$, Fig.~\ref{fig4}(b) illustrates that the LSR decreases towards $0.386$
with the increase of the system size. We also plot the normalized localization length with respect to $L_y$ for different energies, the fall
of which clearly suggests
that the states are localized. These indicators show that all the states are
localized.
Similarly, Fig.~\ref{fig4}(c) indicates that all the states are localized in the gapless HOTAI phase.
But in the Griffiths regime, all the states are localized except at $E=0$
where the LSR remains unchanged as the system size is increased [see Fig.~\ref{fig4}(d)].

\begin{figure}[t]
	\includegraphics[width=3.4in]{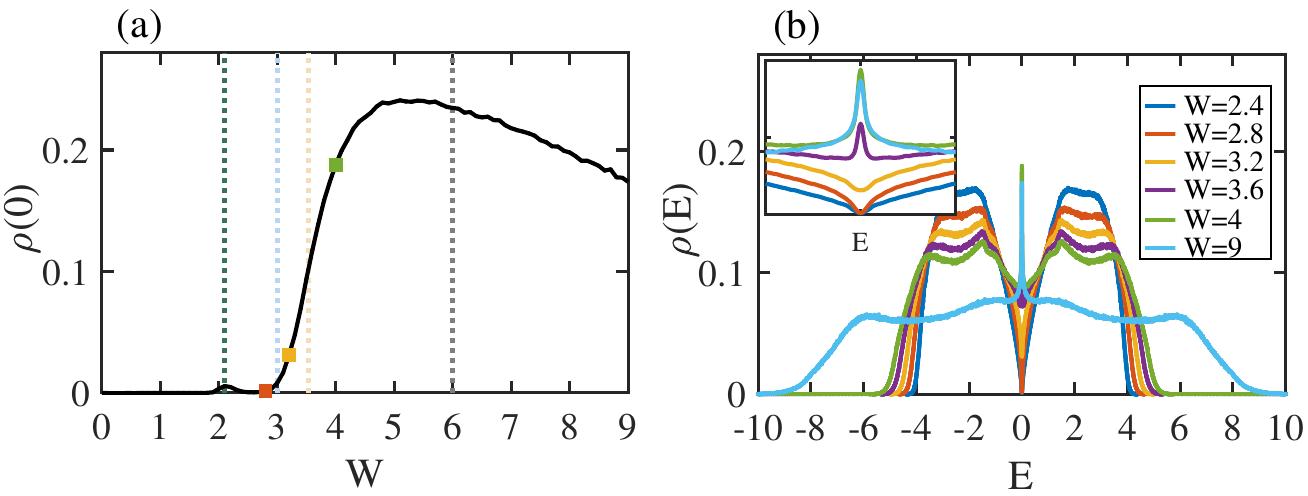}
	\caption{(a) The DOS at zero energy $\rho(0)$ versus the disorder strength $W$
  calculated by the kernel polynomial method (KPM) for the system size $L=100$ and the expansion order $N_c=2^{13}$.
  (b) The DOS $\rho(E)$ versus $E$ for different $W$ with the zoomed-in view around zero energy in the inset.
  The vertical dashed lines separate different phases.
Here $m_x=m_y=1.1$.
	}
	\label{fig5}
\end{figure}

We also compute the density of states (DOS) at $E=0$ with respect to the disorder strength $W$
as shown in Fig.~\ref{fig5}(a).
The DOS is defined as $\rho(E)=\sum_{\bf r}\rho(E,{\bf r})/(4L_x L_y)$, which is normalized to one, i.e., $\int dE\rho(E)=1$.
Here
$\rho(E,{\bf r})=[\sum_{j}\delta(E-E_j)\sum_{\nu=1}^{4}|\Psi_{E_j,{\bf r}\nu}|^2]$ describes the LDOS,
where $\Psi_{E_i,{\bf r}\nu}$ denotes the spatial
eigenstate of the system with periodic boundaries corresponding to the eigenenergy $E_i$, and $[\cdots]$
denotes the average over different samples.
The DOS rises to the maximum in the multifractal phase and then fall in the trivial-II
phase. Specifically, we see the development of a
very narrow peak of the DOS at $E=0$ in this regime [Fig.~\ref{fig5}(b)].

\section{Self-consistent Born approximation}
\label{sec5}

We now explain the disorder induced quadrupole topological insulator based on the self-consistent Born approximation (SCBA)~\cite{Beenakker2009PRL}.
As introduced in Sec.~\ref{sec2}, we consider a disordered system by adding the following random
intra-cell hopping terms at each unit cell ${\bf r}$
\begin{equation}
\begin{aligned}
V({\bf r})&=W\left(
             \begin{array}{cccc}
               0 & -iV_3({\bf r}) & -iV_1({\bf r}) & 0 \\
               iV_3({\bf r}) & 0 & 0 & iV_2({\bf r}) \\
               iV_1({\bf r}) & 0 & 0 & -iV_4({\bf r}) \\
               0 & -iV_2({\bf r}) & iV_4({\bf r}) & 0 \\
             \end{array}
           \right) \\
&=W\left[V_1'({\bf r})U_1 + V_2'({\bf r})U_2 + V_3'({\bf r})U_3 +V_4'({\bf r})U_4\right],
\end{aligned}
\end{equation}
where $V_1'=(V_1+V_2)/2$, $V_2'=(V_1-V_2)/2$, $V_3'=(V_3+V_4)/2$, $V_4'=(V_3-V_4)/2$,
and $U_1=\sigma_y\otimes\sigma_z$, $U_2=\sigma_y\otimes\sigma_0$, $U_3=\sigma_0\otimes\sigma_y$, $U_4=\sigma_z\otimes\sigma_y$.
Here we have changed the notation
in Sec.~\ref{sec2} by
$V^x\rightarrow V_1$, $\bar{V}^x \rightarrow V_2$, $V^y\rightarrow V_3$ and $\bar{V}^y \rightarrow V_4$ for convenience.
Since we are interested in disorder without correlations, we require
\begin{eqnarray}
\langle V_i'({\bf R}) \rangle&=&0   \\
\langle V_i'({\bf R}_1) V_j'({\bf R}_2) \rangle &=&\frac{1}{24}
\delta_{ij}\delta_{{\bf R}_1{\bf R}_2}
\end{eqnarray}
for $i,j=1,2,3,4$
with $\langle \cdots \rangle$ denoting the average over disorder ensembles.

Based on the self-consistent Born approximation, the effective
Hamiltonian at $E=0$ is given by $H_{\textrm{eff}}({\bf k})=H_0({\bf k})+\Sigma(E=0)$ where the self-energy $\Sigma$ in the presence of disorder can be calculated through
the following self-consistent equation
\begin{equation}\label{SE}
\Sigma(E)=\frac{W^2}{96\pi^2}\int_{BZ} d^2{\bf k} \sum_{n=1}^{4} U_n G U_n,
\end{equation}
where $G=[(E+i0^+)I-H_0({\bf k})-\Sigma(E)]^{-1}$.
At energy $E=0$, we find numerically that the self-energy can be expanded as
\begin{equation}
\Sigma=i\Sigma_0 I + \Sigma_x \sigma_y\otimes\sigma_z + \Sigma_y \sigma_0\otimes\sigma_y,
\end{equation}
with $\Sigma_0,\Sigma_x,\Sigma_y$ being real numbers.
It is clear to see that
the topological masses $m_x$ and $m_y$ associated with topological properties are renormalized by disorder to new values
\begin{align}
m_x^\prime &= m_x + \Sigma_x, \\
m_y^\prime &= m_y + \Sigma_y.
\end{align}

Based on Eq.~(\ref{SE}), we first approximate the self-energy $\Sigma$ by taking $\Sigma=0$
in the right-hand side of the equation, yielding
\begin{equation}
\Sigma_\nu = -\frac{W^2}{48\pi^2}\iint_{BZ} d{\bf k} I_\nu,
\end{equation}
where
\begin{eqnarray}
I_\nu &=&\frac{m_\nu+\sin(k_\nu)}{F({\bf k})} \\
F({\bf k}) &=& 2+\sum_{\nu=x,y}[m_\nu^2+2m_\nu\sin(k_\nu)]
\end{eqnarray}
with $\nu=x,y$.
When $m_x>1$ and $m_y>1$, both $\Sigma_x$ and $\Sigma_y$ are negative due to
the positive integrands, leading to a topological phase transition when the
disorder strength $W$ is sufficiently large so that $m_x^\prime<1$ and $m_y^\prime<1$.
We also numerically solve the Eq.~(\ref{SE}) self-consistently to determine $\Sigma_x$
and $\Sigma_y$ and plot the results in Fig.~\ref{fig3}(f).
For weak disorder, the results agree very well with the numerical phase
boundary.

\section{Disorder effects on HOTIs}
\label{sec6}

\begin{figure*}[t]
  \includegraphics[width=5.5in]{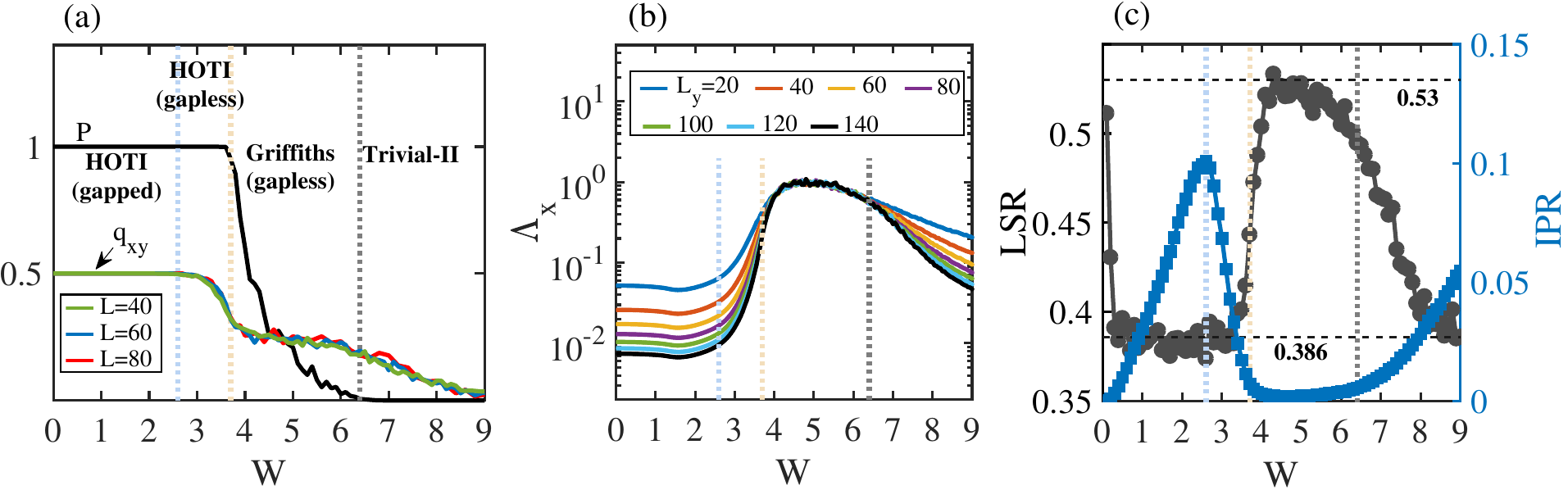}
  \caption{(Color online)
  (a) The phase diagram with respect to $W$, where $P$ and the quadrupole moments
  are displayed. Here the gapped/gapless HOTI,
  the Griffiths phase and the trivial-II phase are observed.
  (b) $\Lambda_x$ at $E=0$ versus $W$ for different $L_y$.
  (c) The LSR and IPR versus $W$ for the eigenstates around zero energy of
  a system with $L=200$.
  The vertical dashed lines separate different phases. Here $m_x=m_y=0.5$.}
\label{fig6}
\end{figure*}

\begin{figure*}[t]
	\includegraphics[width=6.5in]{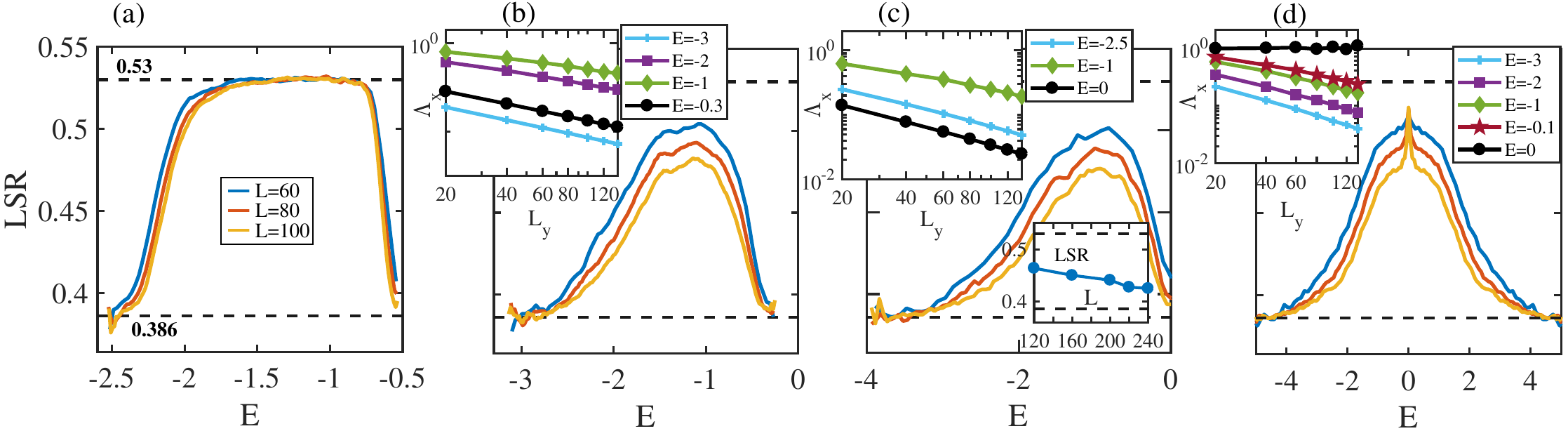}
	\caption{The LSR versus energy $E$ for (a) $W=1$, (b) $W=2$, (c) $W=3.2$ and (d) $W=5$
in gapped HOTI, gapless HOTI and Griffiths phase, respectively.
The top insets in (b), (c) and (d) describe the normalized localization length $\Lambda_x$
with respect to $L_y$ at different energies.
The bottom inset in (c) plots the LSR as a function of system size $L$ at $E=-1$.
Here $m_x=m_y=0.5$.
	}
	\label{fig7}
\end{figure*}

In this section, we study the effects
of disorder on HOTIs. Specifically, we consider $m_{x}=m_{y}=0.5$
corresponding to a HOTI in the clean limit.
We find that the topological phase is stable against weak disorder as evidenced by
the quantized topological invariant $P$ in Fig.~\ref{fig6}(a).
When the disorder strength becomes sufficiently
strong, it enters into a Griffiths regime with fractional $P$ and finally becomes a trivial phase.
The strong disorder also closes the energy gap when $W>2.6$. In the gapless
HOTI and trivial-II phases, all states are localized, as evidenced by the normalized localization
length, LSR and IPR [see Fig.~\ref{fig6}(b) and (c)]. In
the Griffiths regime, the states at $E=0$ are multifractal and all other states
are localized [see Fig.~\ref{fig6}(b)].
In the disordered gapped HOTI, we find that for weak disorder, the states near the band edge are localized as shown by the LSR around $0.386$ in Fig.~\ref{fig6}(c).
For larger disorder, all the states become localized in this phase.

Figure~\ref{fig7} further plots the LSR with respect to $E$ for four different
disorder strength. When the disorder is weak, e.g., $W=1$, the LSR shows that the states near
the band edge are localized in the gapped HOTI [Fig.~\ref{fig7}(a)]. Yet, when $W=2$, the LSR of all the states
decreases towards $0.386$ with the increase of the system size, reflecting that all the states
are localized.
The localized property is also signalled by the decline of the normalized localization length with
increasing $L_y$ [Fig.~\ref{fig7}(b)].
Similarly, in the gapless HOTI, all the states are localized as shown in Fig.~\ref{fig7}(c).
In this case, the LSR at $E=-1$ decreases towards $0.386$ as the system size is increased,
providing further evidence for localization. In the Griffiths regime, the LSR becomes smaller for larger system sizes except
at $E=0$ where it remains unchanged, suggesting that the states at $E=0$ are multifractal
and all other states are localized. The multifractal property is also reflected by the unchanged
property of the normalized localization length as $L_y$ is increased [Fig.~\ref{fig7}(d)].

\section{Experimental realization}
\label{sec7}

\begin{figure*}[t]
	\includegraphics[width=\textwidth]{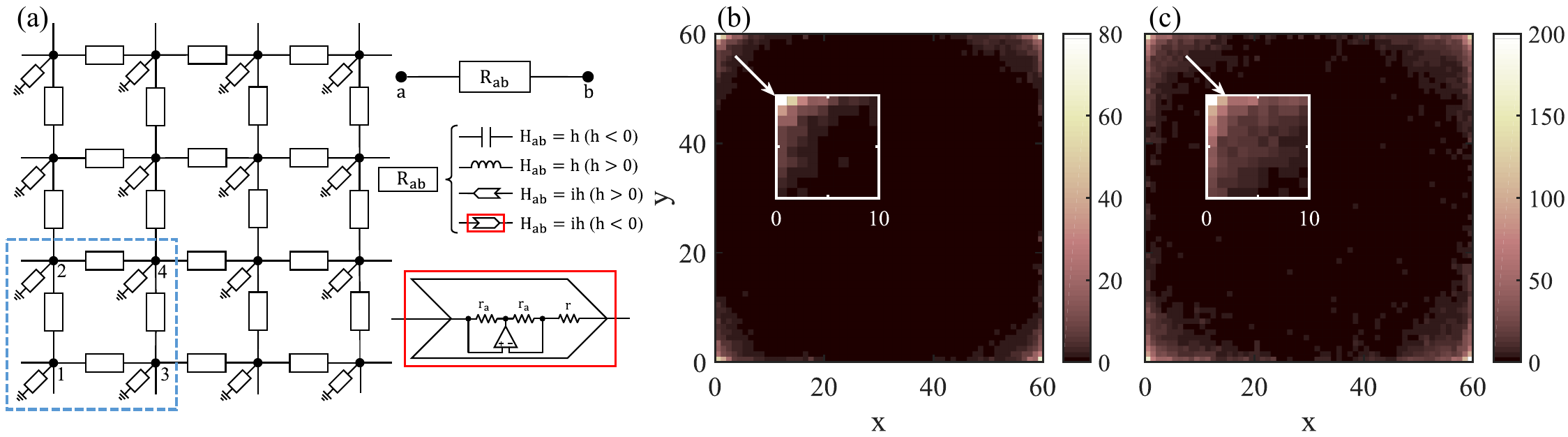}
	\caption{(Color online) (a) Schematics of an electric network for realizing our Hamiltonian (\ref{Hmodel}).
     Here each node in the circuit represents one lattice site in the Hamiltonian,
     and four nodes form a unit cell as shown in the blue dashed box.
     The hopping between two neighboring sites $H_{ab}$ is simulated by the admittance $R_{ab}$ of the electric element connecting them.
     The different values of $H_{ab}$ correspond to different electric elements, including
     capacitors, inductors, or INICs whose
     structure is shown in the red box. Each node $a$ should be grounded through an electric element with
     an appropriate admittance $R_a$.
     (b),(c) The averaged magnitude of the impedance $|Z(x,y)|$ of each unit cell under open boundary conditions
     for $W=2.6$ and $W=3.4$, respectively. The insets show the zoomed-in view of $|Z(x,y)|$ around one corner.
     Here $m_x=m_y=1.1$.
	}
	\label{fig8}
\end{figure*}

The BBH model has been experimentally realized in several metamaterials, such as
microwave, phononic, photonic and topolectrical circuit systems~\cite{Huber2018Nature,Bahl2018Nature,Thomale2018NP,Hafezi2019NP}.
In fact, some systems, such as silicon ring resonators~\cite{Hafezi2019NP}, have demonstrated the robustness
of zero-energy corner modes to certain disorders.
The HOTAI can be easily realized in these systems when the off-diagonal hopping disorder is considered in the experimentally
realized BBH model.
The BBH model has also been implemented in topolectrical circuits,
and zero-energy corner modes are probed by measuring two-point impedances~\cite{Thomale2018NP}.
One can involve disorder in the system by tuning the capacitance of capacitors and inductance of inductors
to realize the HOTAI, as we have proved that this model is equivalent to our model in topological and
localization properties (see Appendix B).
In the following, we discuss in detail an experimental scheme to realize the Hamiltonian (\ref{Hmodel})
using topolectrical circuits and show that the HOTAI phase can be detected by two-point impedance measurements.

Let us consider an electric network composed of different nodes and electric element connecting nodes,
as shown in Fig.~\ref{fig8}(a).
We denote the input current and voltage of each node $a$ by $I_a$ and $V_a$, respectively.
According to Kirchhoff's law, the circuit at a frequency of $\omega$ should satisfy the relation
\begin{equation}
I_a(\omega) = \sum_b R_{ab} (V_a(\omega)-V_b(\omega)) + R_{a} V_a(\omega),
\end{equation}
where $R_{ab}$ is the admittance of the corresponding electric element between the node $a$ and $b$,
and $R_a$ is the admittance of the electric element between the node $a$ and the ground.
We can rewrite the above equation into a compact form as
\begin{equation}
{\bf I}(\omega)=J(\omega) {\bf V}(\omega),
\end{equation}
where ${\bf I}$ and ${\bf V}$ are $N$-component column vectors with components $I_a$ and $V_a$ for $N$ nodes, respectively.
Here the matrix $J$ is the circuit Laplacian. Then we can simulate our Hamiltonian $H$ with
the Laplacian $J(\omega)$ at a proper frequency $\omega$ through
\begin{equation}
J(\omega)=iH.
\end{equation}
Each node in the circuit represents one lattice site in our Hamiltonian,
and each electric element linking two nodes represents the corresponding hopping between the sites,
which can be either a capacitor, or an inductor,
or a negative impedance converter with current inversion (INIC).
For two nodes in the circuit, the electric element between them
is determined according to the corresponding matrix element $H_{ab}$ between the site $a$ and $b$
in our Hamiltonian, as illustrated in Fig.~\ref{fig8}(a).
Specifically, for two neighboring sites in our Hamiltonian,
if $H_{ab}$ is a positive (negative) real number, the electric element between $a$ and $b$ should be an inductor (a capacitor)
with inductance (capacitance) $\frac{1}{\omega H_{ab}}$ ($-\frac{H_{ab}}{\omega}$).
For the case that $H_{ab}$ is an imaginary number, we should connect the two nodes using an INIC with resistance $\frac{1}{|H_{ab}|}$
and proper direction.
In addition, we connect every node with the ground by appropriate electric elements to eliminate the extra diagonal terms in the Laplacian.

Similar to the experimental work~\cite{Thomale2018NP}, we utilize the two-point impedance measurement
in the circuit to characterize the zero-energy corner modes in the HOTAI phase.
The two-point impedance between node $a$ and $b$ is defined as
\begin{equation}
Z_{ab}=\sum_{n}\frac{|\psi_{n,a}-\psi_{n,b}|^2}{j_n},
\end{equation}
where $\psi_{n,a}$ is the component for the node $a$ of the $n$th eigenvector of $J$
with eigenvalue $j_n$.
We define the impedance of each unit cell $Z(x,y)$ as the average two-point impedance between nearest-neighbor
nodes within each unit cell as
\begin{equation}
Z(x,y)=\frac{Z_{12}(x,y)+Z_{24}(x,y)+Z_{43}(x,y)+Z_{31}(x,y)}{4},
\end{equation}
where $Z_{ij}(x,y)$ denotes the two-point impedance between the $i$th node and $j$th node of the unit cell $(x,y)$.
Fig.~\ref{fig8}(b) and (c) plot the magnitude of $Z(x,y)$ averaged over $400$ random samples under open boundary conditions
for two values of $W$ within HOTAI regime for $m_x=m_y=1.1$,
which clearly show the impedance resonance near the corners
corresponding to the presence of zero-energy corner modes of the Hamiltonian.

\section{Conclusion}
\label{sec8}

In summary, we have discovered the HOTAI in a 2D disordered system
with chiral symmetry. Specifically, we show that a topologically trivial
phase can transition into a quadrupole topological phase, when disorder
is added.
We find gapped and gapless HOTAIs and a Griffiths regime. In the gapless HOTAI,
all the states are localized, while in the Griffiths regime,
the states at zero energy are multifractal and other states are localized.
The Griffiths regime corresponds to a critical regime between two localized phases: a gapless HOTAI and a trivial phase.
We also propose an experimental scheme with topolectrical circuits to realize the HOTAI.
Our results demonstrate that disorder can induce quadrupole topological insulators with
peculiar localization properties from a trivial phase and thus opens a new avenue
for studying the role of disorder in higher-order topological phases.

\begin{acknowledgments}
We thank Y.-L. Tao and N. Dai for helpful discussion. Y.B.Y., K.L. and Y.X.
are supported by the National Natural Science Foundation
of China (11974201), the start-up fund from Tsinghua University and
the National Thousand-Young-Talents Program.
We acknowledge in addition support from the Frontier Science Center for Quantum Information of the Ministry of Education of China, Tsinghua University Initiative Scientific Research Program, and the National key Research and Development Program of China (2016YFA0301902).
\end{acknowledgments}

Note added: Recently,
we became aware of two related work~\cite{Shen2020PRL,Zhang2020arXiv}.



\section*{Appendix A: Generalized $C_4$ symmetry}

\setcounter{equation}{0} \setcounter{figure}{0} \setcounter{table}{0}
\renewcommand{\theequation}{A\arabic{equation}} \renewcommand{\thefigure}{A\arabic{figure}}
\renewcommand{\theHfigure}{A\arabic{figure}}
\renewcommand{\bibnumfmt}[1]{[#1]} \renewcommand{\citenumfont}[1]{#1}

In the clean case, when $m_x=m_y$ and $t_x=t_y$, the Hamiltonian (\ref{Hmodel}) in the main text respects a generalized $C_4$ symmetry,
\begin{equation}
U_{C_4}\hat{H}U_{C_4}^{-1}=\hat{H},
\end{equation}
where
\begin{equation}
U_{C_4}\hat{c}_{\bf r} U_{C_4}^{-1}=S_{\bf r}\hat{c}_{g{\bf r}},~~~
U_{C_4}\hat{c}_{\bf r}^\dagger U_{C_4}^{-1}=\hat{c}_{g{\bf r}}^\dagger S_{\bf r}^{T},
\end{equation}
with
\begin{equation}
S_{\bf r}=\left(
            \begin{array}{cccc}
              0 & 0 & (-1)^y & 0 \\
              -(-1)^y & 0 & 0 & 0 \\
              0 & 0 & 0 & (-1)^y \\
              0 & (-1)^y & 0 & 0 \\
            \end{array}
          \right)
\end{equation}
and $g$ being a $C_4$ rotation operator such that $g{\bf r}=(-y,x)
$.

In momentum space, let us write $\hat{H}=\sum_{\bf k}\hat{c}_{\bf k}^\dagger H({\bf k}) \hat{c}_{\bf k}$ with
$\hat{c}_{\bf k}^\dagger=(\begin{array}{cccc}
                      \hat{c}_{{\bf k} 1}^\dagger  & \hat{c}_{{\bf k} 2}^\dagger & \hat{c}_{{\bf k} 3}^\dagger & \hat{c}_{{\bf k} 4}^\dagger
                     \end{array})
$. The generalized $C_4$ symmetry takes the following form
\begin{equation}
S_1^\dagger H({\bf k}) S_1=H(g{\bf k}^\prime),
\end{equation}
where ${\bf k}^\prime=(k_x,k_y-\pi)$ and
\begin{equation}
S_1=\left(
      \begin{array}{cccc}
        0 & 0 & 1 & 0 \\
        -1 & 0 & 0 & 0 \\
        0 & 0 & 0 & 1 \\
        0 & 1 & 0 & 0 \\
      \end{array}
    \right).
\end{equation}

\section*{Appendix B: Equivalence between our model and the disordered BBH model}

\setcounter{equation}{0} \setcounter{figure}{0} \setcounter{table}{0}
\renewcommand{\theequation}{B\arabic{equation}} \renewcommand{\thefigure}{B\arabic{figure}}
\renewcommand{\theHfigure}{B\arabic{figure}}
\renewcommand{\bibnumfmt}[1]{[#1]} \renewcommand{\citenumfont}[1]{#1}
In this Appendix, we will prove that our model is equivalent to the BBH model in topological
and localization properties. The BBH model reads
\begin{equation}
\hat{\tilde{H}}=\sum_{{\bf r}} \left[ \hat{c}^\dagger_{{\bf r}}\tilde{h}_0\hat{c}_{{\bf r}}
+\left( \hat{c}^\dagger_{{\bf r}}h_x\hat{c}_{{\bf r}+{\bf e}_x}
 +\hat{c}^\dagger_{{\bf r}}h_y\hat{c}_{{\bf r}+{\bf e}_y}+H.c. \right) \right],
\end{equation}
where $\tilde{h}_0$ is a real matrix expressed as
\begin{equation}
\tilde{h}_0=\left(
  \begin{array}{cccc}
    0 & m_{{\bf r}}^y & m_{{\bf r}}^x & 0 \\
    m_{{\bf r}}^y & 0 & 0 & -\bar{m}_{{\bf r}}^x \\
    m_{{\bf r}}^x & 0 & 0 & \bar{m}_{{\bf r}}^y \\
    0 & -\bar{m}_{{\bf r}}^x & \bar{m}_{{\bf r}}^y & 0 \\
  \end{array}
\right).
\end{equation}
This model respects the time-reversal, particle-hole and chiral symmetries.

While the two Hamiltonians have different symmetries, they are closely related by
a local transformation $U_{{\bf r}}=\text{diag}(i^{x+y-1},i^{x+y},i^{x+y},i^{x+y+1})$,
that is, $U_{{\bf r}}^\dagger h_0 U_{{\bf r}}=\tilde{h}_0$,
$U_{{\bf r}}^\dagger h_x U_{{\bf r}+{\bf e}_x}={h}_x$ and $U_{{\bf r}}^\dagger h_y U_{{\bf r}+{\bf e}_y}={h}_y$.
Specifically, one can transform $\hat{H}$ in Eq.~(\ref{Hmodel}) to $\hat{\tilde{H}}$ by
the transformation $\hat{c}_{{\bf r}}\rightarrow U_{{\bf r}}\hat{c}_{{\bf r}}$.
In other words, if $\Psi_{E_i,{\bf r}\nu}$ is a spatial eigenstate of $H$,
then $\tilde{\Psi}_{E_i,{\bf r}\nu}=(-i)^{f_{{\bf r}\nu}}\Psi_{E_i,{\bf r}\nu}$ with $f_{{\bf r} 1}=x+y-1$,
$f_{{\bf r} 2}=f_{{\bf r} 3}=x+y$, and $f_{{\bf r} 4}=x+y+1$ is an eigenstate of $\tilde{H}$ corresponding
to the same energy $E_i$. Here $H$ and $\tilde{H}$ are the first-quantization Hamiltonians
of $\hat{H}$ and $\hat{\tilde{H}}$, respectively.
Therefore, $\hat{H}$ and $\hat{\tilde{H}}$ have the same energy spectrum and density profiles, indicating
identical localization properties that they possess. In addition, this local phase transformation
does not change the topological property, and thus the two Hamiltonians have the same topology.
Under open boundary conditions, the two models are connected by the transformation irrelevant of
the system size. Yet, under periodic boundary conditions,
the transformation works well only when $L_x$ and $L_y$ are integer multiples of four.
For topological property, the two models should be equivalent irrelevant of a system size
given that the topology does not depend on a specific system size. For localization property,
we have also calculated the IPR and LSR of the two Hamiltonians with their sizes being odd
and find similar results, showing that their localization properties are irrelevant of the
parity of a system size.

\section*{Appendix C: Quantization of quadrupole moments by chiral symmetry}

\setcounter{equation}{0} \setcounter{figure}{0} \setcounter{table}{0}
\renewcommand{\theequation}{C\arabic{equation}} \renewcommand{\thefigure}{C\arabic{figure}}
\renewcommand{\theHfigure}{C\arabic{figure}}
\renewcommand{\bibnumfmt}[1]{[#1]} \renewcommand{\citenumfont}[1]{#1}

In this Appendix, we will prove that the quadrupole moment is protected to be quantized by chiral
symmetry and thus can be used as a topological invariant. Note that the quadrupole moment may
not characterize the physical quadrupole moment, we here are only interested in the formula
as a topological invariant. We consider
the quadrupole moment defined by~\cite{Wheeler2018arXiv,Cho2018arXiv}
\begin{equation}
\begin{aligned}
q_{xy}&\equiv\left[\tilde{q}_{xy}-q_{xy}^{(0)}\right] \mod 1 \\
&=\left[\frac{1}{2\pi}\mathrm{Im}\log \langle \Psi_G| e^{2\pi i\hat{Q}_{xy}} |\Psi_G\rangle
-q_{xy}^{(0)}\right] \mod 1,
\end{aligned}
\end{equation}
where $\hat{Q}_{xy}=\sum_{j=1}^{n_c}\hat{x}_{j}\hat{y}_{j}/(L_x L_y)$ with $\hat{x}_{j}$ ($\hat{y}_{j}$)
denoting the $x$-position ($y$-position) operator for electron $j$ with $n_c=2L_x L_y$ (the number of occupied states in our model)
at half filling,
and
$|\Psi_G\rangle$ is the many-body ground state of electrons in the system.
Here $\tilde{q}_{xy}=
\frac{1}{2\pi}\mathrm{Im}[\log \langle \Psi_G| e^{2\pi i\hat{Q}_{xy}} |\Psi_G\rangle ]$
is the contribution from occupied electrons,
and $q_{xy}^{(0)}=\frac{1}{2}\sum_{j=1}^{n_a} x_j y_j/(L_x L_y)$ is the contribution from the background positive charge distribution where $(x_j,y_j)$ denotes the position of the $j$th atomic orbital.
Here, $n_a$ is the total number of atomic orbitals so that the single-particle Hamiltonian is a $n_a\times n_a$ matrix.
At half filling, $n_a=2 n_c$.

Let us write the many-body wave function of occupied electrons in real space representation as
\begin{align}
&\Psi_G({\bf r}_1\nu_1,{\bf r}_2\nu_2,\cdots, {\bf r}_{n_c}\nu_{n_c}) \nonumber \\
=&\frac{1}{\sqrt{n_c!}}
\left|
  \begin{array}{cccc}
    \psi_1({\bf r}_1 \nu_1) & \psi_2({\bf r}_1 \nu_1) & \cdots & \psi_{n_c}({\bf r}_1 \nu_1) \\
    \psi_1({\bf r}_2 \nu_2) & \psi_2({\bf r}_2 \nu_2) & \cdots & \psi_{n_c}({\bf r}_2 \nu_2) \\
    \vdots & \vdots & \ddots & \vdots \\
    \psi_1({\bf r}_{n_c} \nu_{n_c}) & \psi_2({\bf r}_{n_c} \nu_{n_c}) & \cdots & \psi_{n_c}({\bf r}_{n_c} \nu_{n_c}) \\
  \end{array}
\right|,
\end{align}
where $\psi_n$ represents the $n$th occupied eigenstate of a first-quantization Hamiltonian.
Then, the quadrupole moment of occupied electrons $\tilde{q}_{xy}$
can be evaluated through
\begin{equation}
\tilde{q}_{xy}=\frac{1}{2\pi}\mathrm{Im} \log \langle \Psi_G|\tilde{\Psi}_G\rangle ,
\end{equation}
where
\begin{equation}
\langle \Psi_G|\tilde{\Psi}_G\rangle=
\left|
  \begin{array}{cccc}
    \langle \psi_1|\tilde{\psi}_1 \rangle & \langle \psi_1|\tilde{\psi}_2\rangle & \cdots & \langle \psi_1|\tilde{\psi}_{n_c}\rangle \\
    \langle \psi_2|\tilde{\psi}_1 \rangle & \langle \psi_2|\tilde{\psi}_2\rangle & \cdots & \langle \psi_2|\tilde{\psi}_{n_c}\rangle \\
    \vdots & \vdots & \ddots & \vdots \\
    \langle \psi_{n_c}|\tilde{\psi}_1\rangle & \langle \psi_{n_c}|\tilde{\psi}_2\rangle & \cdots & \langle \psi_{n_c}|\tilde{\psi}_{2}\rangle \\
  \end{array}
\right|,
\end{equation}
$\tilde{\psi}_n({\bf r}\nu)=e^{i2\pi x y/(L_x L_y)}\psi_n({\bf r}\nu)$
and $\langle \psi_m|\tilde{\psi}_n\rangle=\sum_{{\bf r}\alpha}\psi_m^*({\bf r}\alpha)\tilde{\psi}_n({\bf r}\alpha)$.
Let us define
$U_{o}=\left(|\psi_1 \rangle,|\psi_2 \rangle,\cdots,|\psi_{n_c} \rangle\right)$
which is a $n_a\times n_c$ matrix representing the occupied states of electrons.
Then we can express the quadrupole moment of occupied electrons as
\begin{equation}
\tilde{q}_{xy}=\frac{1}{2\pi}\mathrm{Im}\log \det (U_{o}^\dagger \hat{D} U_{o}),
\end{equation}
where we define a $n_a\times n_a$ diagonal matrix $\hat{D}=\text{diag}\{e^{2\pi i x_j y_j/(L_x L_y)}\}_{j=1}^{n_a}$
with $(x_{j},y_{j})$ denoting the position of $j$-th atomic orbital.

For a generic Hamiltonian in real space $H$ with chiral (sublattice) symmetry, ${\Pi} H {\Pi}^{-1}=-H$,
if $|\psi_n\rangle$ is an eigenstate of $H$ corresponding to energy $E_n$, ${\Pi}|\psi_n\rangle$ is also an eigenstate
of $H$ with energy $-E_n$, corresponding to an unoccupied state. The set
$\{{\Pi}|\psi_1\rangle,{\Pi}|\psi_2\rangle,\cdots, {\Pi}|\psi_{n_c}\rangle\}$ therefore constitutes the unoccupied states.
We then define $U_{u}=\left({\Pi}|\psi_1\rangle,{\Pi}|\psi_2\rangle,\cdots, {\Pi}|\psi_{n_c}\rangle\right)={\Pi}U_{o}$
representing the unoccupied states of electrons.
The quadrupole moment $\tilde{q}_{xy}^{u}$ for the unoccupied states is
\begin{align}
\tilde{q}_{xy}^{u}&=\frac{1}{2\pi}\mathrm{Im}\log \det (U_{u}^\dagger \hat{D} U_{u}) \\
&=\frac{1}{2\pi}\mathrm{Im}\log \det (U_{o}^\dagger \Pi^\dagger \hat{D} \Pi U_{o}).
\end{align}
Clearly, $\hat{D}$ commutes with the chiral (sublattice) symmetry transformation $\Pi$, i.e., $\left[\hat{D},\Pi\right]=0$,
\begin{equation}
\tilde{q}_{xy}^{u}=\frac{1}{2\pi}\mathrm{Im}\log \det (U_{o}^\dagger \hat{D} U_{o})
=\tilde{q}_{xy}.
\end{equation}
Let us define $q_{xy}^{u}\equiv \left[\tilde{q}_{xy}^{u}-q_{xy}^{(0)}\right] \mod 1$.
Then we will have
\begin{equation}
q_{xy}^{u}= \left[\tilde{q}_{xy}-q_{xy}^{(0)}\right] \mod 1 =q_{xy}.
\end{equation}

Next we will prove that $q_{xy}+q_{xy}^{u}=0 \mod 1$, i.e., $\tilde{q}_{xy}+\tilde{q}_{xy}^{u}-2q_{xy}^{(0)} = 0 \mod 1$.
\begin{proof}
We define a unitary matrix $U_{t}=\left(U_{o},U_{u}\right)$.
It can be easily seen that
\begin{align}
2q_{xy}^{(0)}&=\sum_{j=1}^{n_a} x_j y_j/(L_x L_y) \\
&=\frac{1}{2\pi}\mathrm{Im}\log \det \hat{D} \mod 1 \\
&=\frac{1}{2\pi}\mathrm{Im}\log \det (U_{t}^\dagger \hat{D} U_{t}) \mod 1.
\end{align}
Then we have the following relations
\begin{eqnarray}
&&2\pi (-\tilde{q}_{xy}-\tilde{q}_{xy}^{u}+2q_{xy}^{(0)}) \nonumber \\
&=&-\mathrm{Im}\log \det (U_{o}^\dagger \hat{D} U_{o}) - \mathrm{Im}\log \det (U_{u}^\dagger \hat{D} U_{u})+ \nonumber \\
&&\mathrm{Im}\log \det (U_{t}^\dagger \hat{D} U_{t}) \nonumber \\
&=&\mathrm{Im}\log \det (U_{o}^\dagger \hat{D}^\dagger U_{o}) + \mathrm{Im}\log \det (U_{u}^\dagger \hat{D}^\dagger U_{u})+ \nonumber \\
&&\mathrm{Im}\log \det
\begin{pmatrix}
  U_{o}^\dagger D U_{o} & U_{o}^\dagger \hat{D} U_{u} \\
  U_{u}^\dagger D U_{o} & U_{u}^\dagger \hat{D} U_{u} \\
\end{pmatrix} \nonumber \\
&=&\mathrm{Im}\log \det
\begin{pmatrix}
  U_{o}^\dagger \hat{D}^\dagger U_{o} & U_{o}^\dagger \hat{D}^\dagger U_{u} \\
  0 & U_{u}^\dagger \hat{D}^\dagger U_{u} \\
\end{pmatrix}+ \nonumber \\
&&\mathrm{Im}\log \det
\begin{pmatrix}
  U_{o}^\dagger \hat{D} U_{o} & U_{o}^\dagger \hat{D} U_{u} \\
  U_{u}^\dagger \hat{D} U_{o} & U_{u}^\dagger \hat{D} U_{u} \\
\end{pmatrix} \nonumber \\
&=&\mathrm{Im}\log \det \left[
\begin{pmatrix}
  U_{o}^\dagger \hat{D}^\dagger U_{o} & U_{o}^\dagger \hat{D}^\dagger U_{u} \\
  0 & U_{u}^\dagger \hat{D}^\dagger U_{u} \\
\end{pmatrix}
\begin{pmatrix}
  U_{o}^\dagger \hat{D} U_{o} & U_{o}^\dagger \hat{D} U_{u} \\
  U_{u}^\dagger \hat{D} U_{o} & U_{u}^\dagger \hat{D} U_{u} \\
\end{pmatrix} \right] \nonumber \\
&=&\mathrm{Im}\log \det
\begin{pmatrix}
  \mathds{1} & 0 \\
  U_{u}^\dagger \hat{D}^\dagger U_{u} U_{u}^\dagger \hat{D} U_{o} &
  U_{u}^\dagger \hat{D}^\dagger U_{u} U_{u}^\dagger \hat{D} U_{u} \\
\end{pmatrix} \nonumber \\
&=&\mathrm{Im}\log \det (U_{u}^\dagger \hat{D}^\dagger U_{u} U_{u}^\dagger \hat{D} U_{u}) \nonumber \\
&=&\mathrm{Im}\log \det (U_{u}^\dagger \hat{D} U_{u})^\dagger + \mathrm{Im}\log \det (U_{u}^\dagger \hat{D} U_{u}) \nonumber \\
&=&0 \mod 2\pi.
\end{eqnarray}
In the derivation, we have utilized the orthonormal properties
$U_{o}^\dagger U_{o}=U_{u}^\dagger U_{u}=\mathds{1}$, $U_{o}^\dagger U_{u}=0$ and $U_{o}U_{o}^\dagger + U_{u}U_{u}^\dagger = \mathds{1}$.
\end{proof}

Therefore, we have the following relation
\begin{equation}
q_{xy}+q_{xy}^{u}=0 \mod 1.
\end{equation}
Combined with the relation that $q_{xy}^{u}=q_{xy}$, we get the conclusion that $2q_{xy}=0 \mod 1$, namely,
$q_{xy}$ is quantized to $0$ or $0.5$ up to an integer. The result is consistent with our numerical results
where all disorder configurations exhibit quantized quadrupole moments. 
We note that this proof remains valid when we replace $\hat{Q}_{xy}$ and $q_{xy}^{(0)}$ in the definition of $q_{xy}$ with $\hat{Q}_{f}=\sum_{j=1}^{n_{c}} f(\hat{x}_{j},\hat{y}_{j})/(L_{x} L_{y})$ and $q_{f}^{(0)}=\frac{1}{2}\sum_{j=1}^{n_a} f(x_{j},y_{j})/(L_{x} L_{y})$, respectively where $f(x,y)$ is a general function so that the newly defined quantity is also quantized by chiral symmetry like the quadrupole moment.
In addition, one can use the same procedure to
prove the quantization of the octupole moment in 3D protected by chiral symmetry.

\begin{figure}[t]
	\includegraphics[width=2.4in]{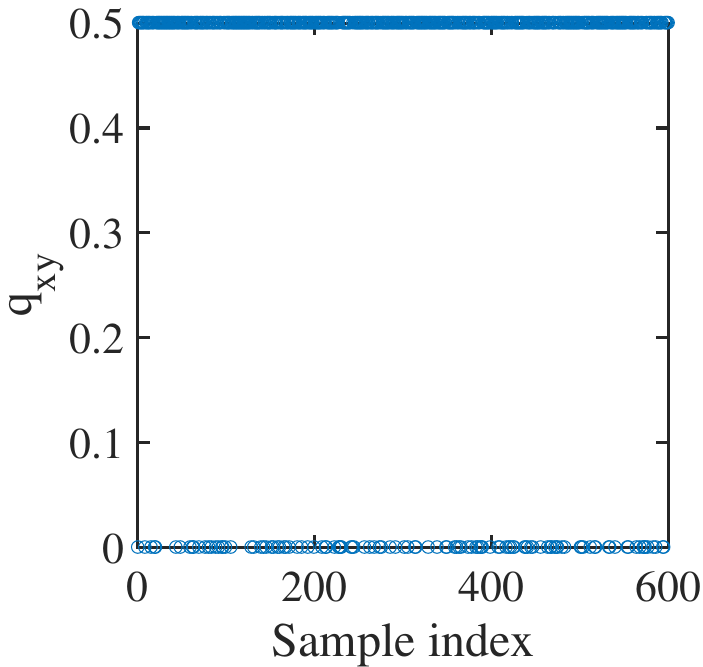}
	\caption{(Color online) The quadrupole moment $q_{xy}$ for different disorder configurations when $W=2.6$, showing that $q_{xy}=0.5$
for most disorder samples and $q_{xy}=0$ for others.
Here $L_x=L_y=80$ and $m_x=m_y=1.1$.
	}
	\label{figC1}
\end{figure}

We note that while we have proved that the quadrupole moment is quantized to either $0$ or $0.5$ for each disorder configuration
protected by chiral symmetry, for a disorder system, we need to consider many distinct samples and perform the average of the quadrupole moment over these samples. In this case, the averaged quadrupole moment may not be quantized since for some samples the quadruple moments are equal to $0.5$ and for others they are equal to $0$ when a system size is not large as shown in Fig.~\ref{figC1}.

It is worth mentioning that Ref.~\cite{Agarwala2020PRR} has found that
quadrupole topological insulators with quantized quadrupole moments can still exist even in amorphous systems without crystalline symmetries.
We now can understand that the quantized quadrupole moment found in Ref.~\cite{Agarwala2020PRR} is protected by chiral symmetry.

\section*{Appendix D: Effective boundary Hamiltonian}

\setcounter{equation}{0} \setcounter{figure}{0} \setcounter{table}{0}
\renewcommand{\theequation}{D\arabic{equation}} \renewcommand{\thefigure}{D\arabic{figure}}
\renewcommand{\theHfigure}{D\arabic{figure}}
\renewcommand{\bibnumfmt}[1]{[#1]} \renewcommand{\citenumfont}[1]{#1}

In this section, we follow the transfer matrix method introduced in Ref.~\cite{Oppen2017PRB} to derive the effective boundary Hamiltonian of our system in the clean case. We will show that the effective boundary Hamiltonian at the $y$-normal ($x$-normal)
edges are proportional to $H_x(k_x,m_x)$ [$H_y(k_y,m_y)$] up to a nonzero factor, implying that the higher-order topology can
be characterized by the topological invariant $P$ introduced in the main text.

Specifically, let us write the Hamiltonian as $\hat{H}=\hat{c}^\dagger H\hat{c}$ where
$\hat{c}^\dagger=(\begin{array}{cccc}
                   \hat{c}_1^\dagger & \hat{c}_2^\dagger & \cdots & \hat{c}_{L_y}^\dagger
                 \end{array})
$ with the index $j$ denoting the $j$th layer consisting of sites along $x$ and $H$ reads
\begin{equation}
H=\left(
    \begin{array}{ccccccc}
      h_1 & V_1^\dagger & 0 & 0 & 0 & \cdots & 0 \\
      V_1 & h_2 & V_2^\dagger & 0 & 0 & \cdots & 0 \\
      0 & V_2 & h_3 & V_3^\dagger & 0 & \cdots & 0 \\
      0 & 0 & V_3 & h_4 & V_4^\dagger & \cdots & 0 \\
      \vdots & \vdots & \vdots & \ddots & \ddots & \ddots & \vdots  \\
      0 & 0 & 0 & \cdots & V_{2L_y-2} & h_{2L_y-1} & V_{2L_y-1}^\dagger \\
      0 & 0 & 0 & \cdots & 0 & V_{2L_y-1} & h_{2L_y} \\
    \end{array}
  \right)
\end{equation}
with $V_{n}$ denoting the coupling between
the $n$th and $(n+1)$th layer. In disordered systems, the parameters in $h_n$ and $V_{2n-1}$ describing the intra-cell hopping
are randomly generated.

In the clean case, the system has the translational invariance of period 2 and thus there are two different layers
described by the Hamiltonian $h_1$ and $h_2$, respectively. If we view these two layers as
a unit cell, we use $V_1$ and $V_2$ to describe the intra-cell and inter-cell layer coupling, respectively.
Now $H$ can be simplified as
\begin{equation}
H=\left(
    \begin{array}{ccccccc}
      h_1 & V_1^\dagger & 0 & 0 & 0 & \cdots & 0 \\
      V_1 & h_2 & V_2^\dagger & 0 & 0 & \cdots & 0 \\
      0 & V_2 & h_1 & V_1^\dagger & 0 & \cdots & 0 \\
      0 & 0 & V_1 & h_2 & V_2^\dagger & \cdots & 0 \\
      \vdots & \vdots & \vdots & \ddots & \ddots & \ddots & \vdots \\
      0 & 0 & 0 & \cdots & V_2 & h_1 & V_1^\dagger \\
      0 & 0 & 0 & \cdots & 0 & V_1 & h_2 \\
    \end{array}
  \right).
\end{equation}

Considering the periodic boundaries along $x$, we write $h_1$, $h_2$, $V_1$ and $V_2$ in momentum space as
$h_1=-h_2=H_x(k_x,m_x)$, $V_1=im_y\sigma_0$ and $V_2=\sigma_0$. When $m_y=0$,
it is clear to see that the effective boundary Hamiltonian is $H_x(k_x,m_x)$.
When $m_y \neq 0$, we obtain the following two transfer matrices at energy $E$
\begin{equation}
\begin{aligned}
&M_1(E)=\left(
      \begin{array}{cc}
        i(E\sigma_0-h_1)/m_y & -i\sigma_0/m_y \\
        \sigma_0 & 0_{2\times 2} \\
      \end{array}
    \right), \\
&M_2(E)=\left(
      \begin{array}{cc}
        (E\sigma_0-h_2) & -im_y\sigma_0 \\
        \sigma_0 & 0_{2\times 2} \\
      \end{array}
    \right),
\end{aligned}
\end{equation}
where the transfer matrices connect the eigenstate in neighboring layers through
\begin{equation}
\begin{aligned}
&\left(
  \begin{array}{c}
    \Psi_{2n} \\
    \Psi_{2n-1} \\
  \end{array}
\right)
=M_1
\left(
  \begin{array}{c}
    \Psi_{2n-1} \\
    \Psi_{2n-2} \\
  \end{array}
\right), \\
&\left(
  \begin{array}{c}
    \Psi_{2n+1} \\
    \Psi_{2n} \\
  \end{array}
\right)
=M_2
\left(
  \begin{array}{c}
    \Psi_{2n} \\
    \Psi_{2n-1} \\
  \end{array}
\right),
\end{aligned}
\end{equation}
with $n\ge 1$ and $\Psi_n$ is the component in the $n$th layer of an eigenstate with the energy $E$.

\begin{figure*}[t]
	\includegraphics[width=\textwidth]{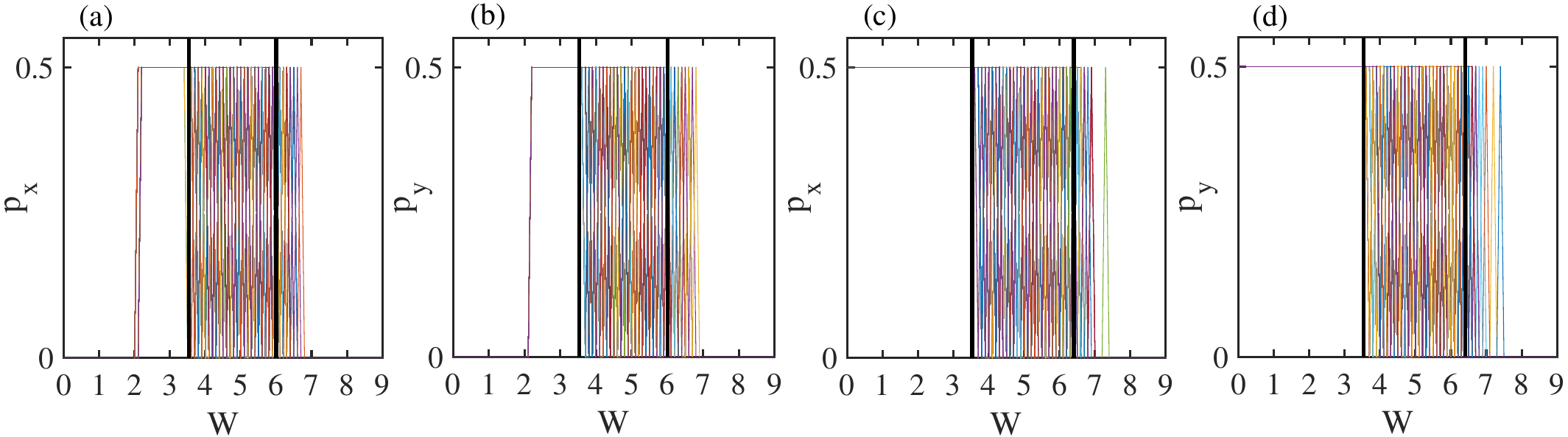}
	\caption{(Color online) The polarization (a) and (c) $p_x$, and (b) and (d) $p_y$ for the $801$th to $1000$th iterations
corresponding to different disorder configurations. The region between two vertical lines
correspond to the Griffiths regime.
In (a-b), $m_x=m_y=1.1$, and in (c-d), $m_x=m_y=0.5$. Here $L_x=L_y=500$.
	}
	\label{figE1}
\end{figure*}

We now define the total transfer matrix at zero energy as
\begin{equation}
T=M_2(E=0)M_1(E=0)=-\frac{i}{m_y}\left(
           \begin{array}{cc}
             A(k_x)\sigma_0 & h_1 \\
             h_1 & \sigma_0 \\
           \end{array}
         \right),
\end{equation}
where $A(k_x)=1+m_x^2+m_y^2+2m_x\sin k_x$.
This matrix can be reduced to a diagonal block form through an elementary interchange transformation,
\begin{equation}
S_{24}TS_{24}=-\frac{i}{m_y}\left(
                 \begin{array}{cc}
                   H_1 & 0_{2\times 2} \\
                   0_{2\times 2} & H_2 \\
                 \end{array}
               \right),
\end{equation}
where
\begin{equation}
\begin{aligned}
&S_{24}=\left(
         \begin{array}{cccc}
           1 & 0 & 0 & 0 \\
           0 & 0 & 0 & 1 \\
           0 & 0 & 1 & 0 \\
           0 & 1 & 0 & 0 \\
         \end{array}
       \right), \\
&H_1=\left(
      \begin{array}{cc}
        A(k_x) & -im_x+e^{-ik_x} \\
        im_x+e^{ik_x} & 1 \\
      \end{array}
    \right), \\
&H_2=\left(
      \begin{array}{cc}
        1 & -im_x+e^{-ik_x} \\
        im_x+e^{ik_x} & A(k_x) \\
      \end{array}
    \right).
\end{aligned}
\end{equation}
Evidently, $H_1$ and $H_2$ have the same eigenvalues. Since $T$ is a symplectic matrix,
its eigenvalues show up in pairs as $(\lambda,1/\lambda^*)$.
Suppose that $\lambda_1 m_y$ ($|\lambda_1|>1$) is an eigenvalue of $H_1$, then
\begin{align}
TU&=T\left(
   \begin{array}{cc}
     U_{11} & U_{12} \\
     U_{21} & U_{22} \\
   \end{array}
 \right) \nonumber \\
&=-i
 \left(
   \begin{array}{cc}
     U_{11} & U_{12} \\
     U_{21} & U_{22} \\
   \end{array}
 \right)
 \left(
   \begin{array}{cc}
     \lambda_1 \sigma_0 & 0_{2\times 2} \\
     0_{2\times 2} & \sigma_0/\lambda_1^* \\
   \end{array}
 \right),
\end{align}
where $U$ is made up of eigenvectors of $H_1$ and $H_2$. Then,
the fixed-point boundary Green's function is given by
\begin{equation}
G(E=0)=\lim_{N\rightarrow \infty}G_N(E=0) =U_{21}(U_{11})^{-1}(V_0^\dagger)^{-1},
\end{equation}
where $V_0$ can be chosen as any invertible matrix. By calculating eigenvectors of $H_1$ and $H_2$, we obtain
the effective boundary Hamiltonian along $x$
\begin{equation}
H_{eff}(k_x)=-G(E=0)^{-1}=\frac{g(k_x)}{f(k_x)}H_x(k_x,m_x),
\end{equation}
where $g(k_x)=-(m_y\lambda_1-1)$ and $f(k_x)=1+m_x^2+2m_x\sin k_x\ge(1-|m_x|)^2>0$ ($m_x=1$ is not considered as it corresponds to a phase boundary). Let us further prove that $g(k_x)<0$ for all $k_x$.
Suppose $m_y>0$, then $\lambda_1=\frac{1}{2m_y}[A(k_x)+1+\sqrt{(A(k_x)+1)^2-4m_y^2}]$, we have
\begin{eqnarray}
2g(k_x)&=&-A(k_x)+1-\sqrt{(A(k_x)+1)^2-4m_y^2} \nonumber \\
&\le& -(m_y^2+m_x^2-2|m_x|) \nonumber \\
&&-\sqrt{[1+m_y^2+(1-m_x)^2]-4m_y^2} \nonumber \\
&<&-(m_y^2+m_x^2-2|m_x|) \nonumber \\
&&-|1-m_y^2+(1-|m_x|)^2|,
\end{eqnarray}
where we have used $A(k_x)\ge (1-|m_x|)^2+m_y^2$. If $1-m_y^2+(1-|m_x|)^2>0$, then $g(k_x)< -(1-|m_x|)^2<0$;
otherwise, we have $m_y^2+m_x^2-2m_x>2(1-|m_x|)^2>0$, giving $g(k_x)<0$.
Similarly,
\begin{equation}
H_{eff}(k_y)=\frac{\bar {g}(k_y)}{\bar {f}(k_y)}H_y(k_y,m_y),
\end{equation}
where $\bar {g}(k_y)=-(m_x \lambda_2 +1)$ and $\bar {f}(k_y)=1+m_y^2+2m_y\sin k_y$ with $\lambda_2=-\frac{1}{2m_x}[B(k_y)+1+\sqrt{(B(k_y)+1)^2 - 4m_x^2}]$
and $B(k_y)=1+m_x^2+m_y^2 + 2m_y \sin k_y$.

Evidently, the higher-order topological
phase arises when these effective boundary Hamiltonians become topological and thus can be
characterized by the topological invariant $P$.

\section*{Appendix E: Griffiths regime}

\setcounter{equation}{0} \setcounter{figure}{0} \setcounter{table}{0}
\renewcommand{\theequation}{E\arabic{equation}} \renewcommand{\thefigure}{E\arabic{figure}}
\renewcommand{\theHfigure}{E\arabic{figure}}
\renewcommand{\bibnumfmt}[1]{[#1]} \renewcommand{\citenumfont}[1]{#1}

In the main text, we have shown the existence of a Griffiths phase where
topologically nontrivial and trivial samples coexist, leading to
the topological invariant $P$ that is not quantized. In Fig.~\ref{figE1}, we plot
the polarizations in $200$ different iteration steps corresponding to different
sample configurations. We see that in the Griffiths regime, some results
show the polarization of $0.5$ and others zero.

\section*{Appendix F: The finite-size analysis of the LSR and IPR}

\setcounter{equation}{0} \setcounter{figure}{0} \setcounter{table}{0}
\renewcommand{\theequation}{F\arabic{equation}} \renewcommand{\thefigure}{F\arabic{figure}}
\renewcommand{\theHfigure}{F\arabic{figure}}
\renewcommand{\bibnumfmt}[1]{[#1]} \renewcommand{\citenumfont}[1]{#1}

\begin{figure*}[t]
	\includegraphics[width=4.5in]{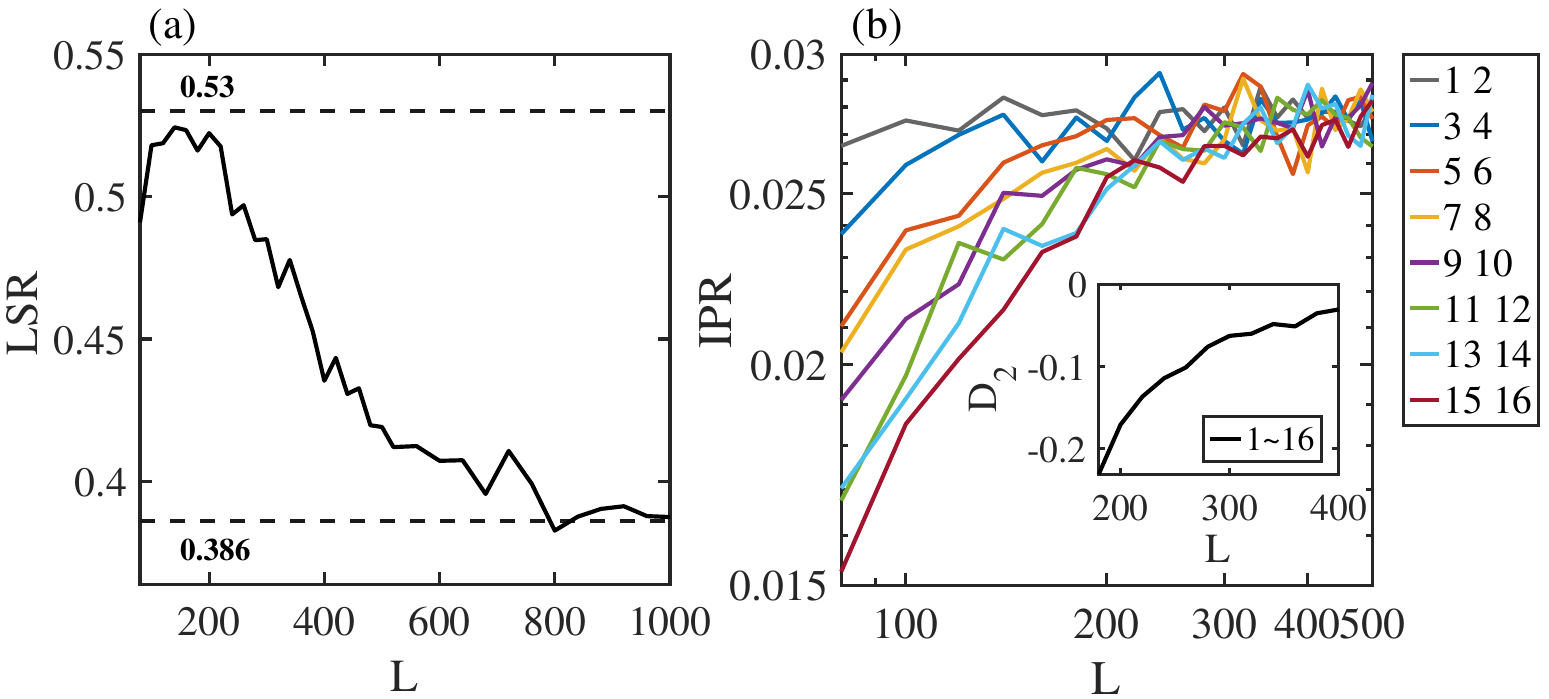}
	\caption{(Color online) (a) The LSR with respect to the system size $L$ for $W=1$ and (b) the IPR with respect to the system size $L$ for $W=3.2$
for different energy levels around zero energy. The inset shows the fractal dimension $D_2$ averaged over these energy levels as a function of $L$.
Here $m_x=m_y=1.1$.
	}
	\label{figF1}
\end{figure*}

In the main text, we have shown that the LSR at the band edge around zero energy in the region around $W=1$ is
close to $0.386$, indicating that the states are localized. Here we further plot the LSR for $W=1$ with respect to
the system size in Fig.~\ref{figF1}(a), illustrating that the LSR approaches $0.386$ as the system size increases.
We have also shown in the main text that for the localized states, the
fractal dimension $D_2$ can take negative values due to finite-size effects. Here we plot the IPR with respect to
the system size in Fig.~\ref{figF1}(b), showing the increase of the IPR with respect to the system size for a system with moderate sizes.
Such an increase gives a negative fractal dimension. Yet, the increase slope declines as the system size is raised,
indicating that $D_2$ approaches zero in the thermodynamic limit.


\end{document}